\documentclass[authoryear,12pt]{article}

\usepackage{natbib}
\usepackage[title]{appendix}
\usepackage[english]{babel}
\usepackage{amsmath}
\usepackage{latexsym}
\usepackage{amssymb}
\usepackage{amsmath}
\usepackage{amsthm}
\usepackage{mathtools}
\usepackage{multirow}
\usepackage{multicol}
\usepackage{float}
\usepackage{graphicx}
\usepackage{booktabs}
\usepackage{hyperref}
\usepackage{array}
\usepackage{hyperref}
\usepackage{enumitem}
\usepackage{colortbl}
\usepackage{subfig}
\usepackage{epsfig}
\usepackage{lscape}
\usepackage[normalem]{ulem}
\usepackage[usenames,dvipsnames]{xcolor}
\usepackage{tikz}
\usetikzlibrary{bayesnet}
\usepackage{bbm}
\usepackage{caption} 
\usepackage{rotating}
\usepackage{algorithm}
\usepackage{algpseudocode}
\usepackage[blocks]{authblk} 
\makeatletter
\newcommand*{\centernot}{%
	\mathpalette\@centernot
}
\def\@centernot#1#2{%
	\mathrel{%
		\rlap{%
			\settowidth\dimen@{$\m@th#1{#2}$}%
			\kern.5\dimen@
			\settowidth\dimen@{$\m@th#1=$}%
			\kern-.5\dimen@
			$\m@th#1\not$%
		}%
		{#2}%
	}%
}
\makeatother

\newcommand{\vertiii}[1]{{\left\vert\kern-0.20ex\left\vert\kern-0.20ex\left\vert #1 
	\right\vert\kern-0.20ex\right\vert\kern-0.20ex\right\vert}}

	\newcommand{\lnorm}[1]{\left\lVert#1\right\rVert}

	\newtheorem{defin}{Definition}

	\newtheorem{theorem}{Theorem}
	\newtheorem{corollary}{Corollary}
	\newtheorem{proposition}{Proposition}

\newcounter{rmk}
\newtheorem{rem}[rmk]{Remark}
\newcounter{example}
\newtheorem{ex}[example]{Example}

\newcommand{\Z}{\mathbb{Z}}
\newcommand{\R}{\mathbb{R}}
\newcommand{\E}{\mathrm{E}}

\newcommand{\I}{\mathbf{I}}
\newcommand{\Hb}{\mathbf{H}}
\newcommand{\X}{\mathbf{X}}

\newcommand{\eb}{\mathbf{e}}
\renewcommand{\P}{\mathrm{P}}
\newcommand{\Fb}{\mathcal{F}}

\newcommand{\Yc}{\mathcal{Y}}
\newcommand{\Lambdab}{\boldsymbol{\Lambda}}
\newcommand{\Vb}{\boldsymbol{V}}
\newcommand{\zerob}{\boldsymbol{0}}
\newcommand{\Zb}{\mathbf{Z}}

\newcommand{\yb}{\mathbf{y}}
\newcommand{\Y}{\mathbf{Y}}
\newcommand{\A}{\mathbf{A}}
\newcommand{\B}{\mathbf{B}}
\newcommand{\G}{\mathbf{G}}
\newcommand{\W}{\mathbf{W}}
\newcommand{\w}{\mathbf{w}}

\newcommand{\kb}{\mathbf{k}}
\newcommand{\fb}{\mathbf{f}}

\newcommand{\lambdab}{\boldsymbol{\lambda}}
\newcommand{\varepsilonb}{\boldsymbol{\varepsilon}}

\newcommand{\nub}{\boldsymbol{\nu}}
\newcommand{\thetab}{\boldsymbol{\theta}}

\newcommand{\pib}{\boldsymbol{\pi}}
\newcommand{\pb}{\boldsymbol{p}}
\newcommand{\etab}{\boldsymbol{\eta}}
\newcommand{\Sigmab}{\boldsymbol{\Sigma}}

\newcommand{\gammab}{\boldsymbol{\gamma}}
\newcommand{\taub}{\boldsymbol{\tau}}
\newcommand{\Tb}{\mathcal{T}}
\newcommand{\Pb}{\mathbf{P}}
\newcommand{\Db}{\mathbf{D}}
\newcommand{\Sb}{\mathbf{S}}
\newcommand{\Jb}{\mathbf{J}}
\newcommand{\ffb}{\mathbf{F}}
\newcommand{\ssb}{\mathbf{s}}
\newcommand{\vecc}{\text{vec}}

\hypersetup{
colorlinks=true,
linkcolor=blue,
filecolor=blue,      
urlcolor= blue,
citecolor=blue,
}

\DeclareMathOperator*{\argmin}{arg\,min}

\DeclareMathOperator*{\argmax}{arg\,max}

\begin{document}

\title{Two-stage weighted least squares estimator of multivariate non-negative observation-driven models
}
\date{} 
\author{Mirko Armillotta \footnote{Email address: mirko.armillotta@uniroma2.it.}}
\affil{Department of Economics and Finance, University of Rome Tor Vergata} 

{
	\makeatletter
	\renewcommand\AB@affilsepx{: \protect\Affilfont}
	\makeatother
	
	
	\makeatletter
	\renewcommand\AB@affilsepx{, \protect\Affilfont}
	\makeatother
	
}

{\let\newpage\relax\maketitle}

\vspace{-0.1cm}



	\begin{abstract}
		\noindent
A novel estimation approach for a general class of semi-parametric multivariate time series
models is introduced where the conditional mean is modeled through parametric functions.
The focus of the estimation is the conditional mean parameter vector for non-negative time series.
Quasi-Maximum Likelihood Estimators (QMLEs) based on the linear exponential family are typically employed for such estimation problems when the true multivariate conditional probability distribution is unknown or too complex. Although QMLEs provide consistent estimates they may be inefficient.
Novel two-stage Multivariate Weighted Least Square Estimators (MWLSEs) are introduced which enjoy the same consistency property as the QMLEs but provide improved efficiency with a suitable choice of the weighting sequence of matrices in the second stage. 
The proposed method  enables a more accurate estimation of model parameters, particularly for data where maximum likelihood estimation is infeasible.
Moreover, consistency and asymptotic normality of MWLSEs are derived, and their efficiency is proved under the correct specification of the weighting sequence.
The estimation performance of QMLE and MWLSE is also compared through simulation experiments and a real data application,  showing the superior accuracy of the proposed methodology. 
\end{abstract}


\section{Introduction} \label{SEC: introduction}

A large variety of statistical models have been proposed to model the conditional mean of time series data. The specification of the model strongly depends on the nature of the time series  variables of the analysis.  An example of models typically employed when the time series variables are continuous and take values on $\R$ are the well-known Autoregressive Moving Average (ARMA) models \citep{box1970time}.  Instead, when variables are non-negative, i.e. their domain is in the non-negative real line (or a subset of it), modelling approaches have been developed. For example, Autoregressive Conditional Duration (ACD) models  \cite[]{engle1998autoregressive} are tailored   for modeling   positive  continuous  processes and   beta autoregressive models \cite[]{rocha2009beta}  are employed for modeling  time series data lying in a bounded interval. With regard to discrete variables,   Integer-valued Autoregressive (INAR) models  \cite[]{al1987, MCK1988} and   Integer-valued GARCH  models (INGARCH) \cite[]{Heinen(2003), fer2006}  have been proposed to model the discrete non-negative count processes. \cite{qaqish2003family},
\cite{russell2005discrete} and \cite{moysiadis2014binary} introduced analogous autoregressive models for binary and categorical time series. Description of further modelling approaches for discrete-valued time series can be found in \cite{armillotta_2022_EJS} and \cite{armillotta2023overview}. Some of the above models have also been extended to the multivariate framework. For example, \cite{hautsch2012econometrics} described a multivariate version of the ACD model.
 \cite{fok2020} introduced the multivariate INGARCH model for vector count time series. A few models for multivariate binary and categorical time series have also been proposed \citep{raftery1985model} and multivariate logistic regressions following the line of \cite{azzalini1994logistic}. This review of modelling approaches is far from being comprehensive. See \cite{joe1997multivariate} and \cite{macdonald1997hidden} for the description of further modelling techniques for non-negative data.

The estimation of the mentioned models can be performed by appealing to Maximum Likelihood Estimator (MLE), which  constitutes a natural approach for the estimation of unknown parameters in both univariate and multivariate parametric models.  However, the MLE lies on parametric assumptions regarding the whole joint distribution of the time series. This constitutes a limiting feature in cases where the interest is only on modeling the conditional means rather than the whole distribution. Moreover, the likelihood function can sometimes present a complex form  and the implementation of the MLE can become cumbersome or unfeasible. For instance, exact likelihood inference of INAR models is well-known to be numerically difficult \cite[]{bu_2008maximum,drost2009efficient,pedeli2015likelihood}. Moreover, these challenges may be exacerbated in the context of multivariate models;
 see \cite{bindai2013multbernoulli} and \cite{fokianos_2021}, among others. In such situations, the use of quasi-likelihood methods becomes attractive.

The   Quasi-MLE (QMLE), proposed by \cite{wedderburn1974quasi}, follows the same philosophy of the MLE but where the object of the optimization is a quasi-likelihood, not necessarily the true distribution of the data. Quasi-likelihoods are typically a member of the one-parameter exponential family in univariate models. In multivariate models the marginal quasi-likelihoods including only the marginal distributions of the data are most often employed. The QMLE is consistent for the true unknown parameters of the model \citep{gourieroux1984pseudo} but it can lead to inefficient estimates for two main reasons: (i) given a parametric function for the conditional mean of the time series, the  conditional variance is implicitly  constrained to be a function of the mean according to the selected distribution within the exponential family; (ii) in multivariate models the marginal QMLE does not account for cross-sectional dependence between the observations.
To tackle the  source of inefficiency (i) for the estimation of the parameters of the conditional mean in univariate time series models,  \cite{francq_2021two} introduce a two-stage Weighted Least Squares Estimator (WLSE) where first-stage estimates are employed to compute a working conditional variance of the process which is then used in the second stage as a weighting sequence for the solution of the weighted least squares problem. WLSEs lead to improved efficiency with respect to QMLEs if the variance function is correctly specified.
This class of univariate two-stage procedures has previously been investigated by \cite{aknouche2012multistage} for ARCH models and has also been recently applied to multiplicative models for count data \citep{aknouche2024multiplicative,weiss2024conditional}.  
 See also \cite{armillotta_gorgi_2023pseudo} for an alternative improved efficiency estimator leveraging simultaneous estimation of conditional mean and variance.
 
 In the multivariate case, similar WLS methods have been considered. See for example \cite{wooldridge1994estimation}, among others.
  Nevertheless, two-stage estimation has been shown to reduce inefficiency only for univariate time series models which does not take into account problem (ii) arising when only marginal distributions are included in multivariate estimation problems.

A novel class  of  two-stage Multivariate Weighted Least Squares Estimator (MWLSEs) for the estimation of the conditional mean of semi-parametric time series models is proposed. In the first stage estimation a weighted least squares solution is obtained for the mean parameters if preliminary information on the weighting sequence is available, otherwise a least squares solution is employed. The first-stage estimates of the unknown parameters are then employed as a basis to compute a working conditional covariance matrix for the process.
This is then used in the second stage as a weighting sequence of matrices for the solution of the multivariate weighted least squares problem. MWLSEs are able to overcome problem (i) since the conditional variances of the process can take a general form and are not required to be correctly specified. Moreover, they can depend on additional parameters. The proposed family of estimators also allows avoiding problem (ii) by taking into account cross-sectional correlations between the observations. Analogously to the specification of the conditional covariances, the correlation matrix is allowed to be misspecified and to depend on extra parameters. The described features show that MWLSEs provide a quite flexible specification of the second moment that can ultimately enhance the efficiency of the estimates with respect to previous QMLE-based approaches. An example of superior performance of MWLSE is provided through extensive numerical results for multivariate integer-valued time series models, which yield evidence of improved estimator efficiency where a moderate or high degree of cross-sectional dependence occurs in the observations. 

Moreover, the MWLSE only requires parametric assumptions on the conditional expectation as the conditional covariance matrix  does not need  to be correctly specified and does not require the specification of the entire conditional probability distribution of the process. In this way, the estimation is robust against possible misspecification of the conditional distribution and simplifies the inference when the full MLE is infeasible or too complex to be specified.  Strong consistency and asymptotic normality of the MWLSEs are established under general conditions, and its efficiency is proved under correct specification of the weighting sequence.

Finally, the practical usefulness of the MWLSE approach is illustrated by means of a real data application concerning binary time series.  In financial time series analysis expected values of  stock returns are typically considered to be unpredictable. However, the direction of price change (positive or negative) can be predicted. Therefore a multivariate binary time series model is proposed for the probability of a positive sign of returns, showing that the two-stage MWLSE improves the estimation capacity of the model with respect to a Bernoulli QMLE. The proposed approach is readily extendable to more general multivariate categorical time series.

The remainder of the paper is organized as follows. Section~\ref{ineff QMLE} discusses the QMLE inefficiency problem in semi-parametric time series models with particular attention to non-negative data. Section~\ref{MWLE} introduces the general framework of  MWLSEs. 
 Asymptotic theory of first-stage and second-stage MWLSEs is also established in this section together with asymptotic efficiency results with respect to QMLEs. In Section~\ref{SEC examples} applications of the established asymptotic theory to specific models of interest are presented under low-level conditions. In
Section~\ref{SEC simulations} an extensive numerical analysis for integer-valued data with the comparison between the MWLSE and alternative quasi-likelihood methods is presented, confirming the improved efficiency performance of MWLSEs. In Section~\ref{SEC: application} an empirical application on IT companies' stock price directions involving binary time series is presented. Finally, Section~\ref{SEC conclusions} concludes. The proofs of all the results are provided in the Appendix.

\section{Inefficiency of QMLE} \label{ineff QMLE}

Consider a multivariate stationary and ergodic process $\{\Y_t\}_{t\in\Z}$  defined in a subset of $\R^d$ with first   conditional moment given by
\begin{equation}
\E(\Y_t|\Fb_{t-1}) = \lambdab(\Y_{t-1}, \Y_{t-2}, \dots; \thetab_0)=\lambdab_t(\thetab_0)\,, \quad t\in \Z
\label{mean equation}
\end{equation}
where $\Z$ is the set of integers, $\Fb_t$ denotes the $\sigma$-field generated by $\{\Y_{s}\,,\,\, s\leq t\}$, $ \lambdab : \R^\infty \times \Theta \to \Lambdab$ is a known 
measurable function in $\Lambdab \subseteq \R^d$, with $\thetab_0$ unknown parameter vector belonging to some compact parameter space $\Theta \subseteq \R^{m}$. 
Model \eqref{mean equation} is  an observation-driven model \citep{cox1981} since the conditional mean parameter purely depends only on past observations.
Sometimes the specification of model \eqref{mean equation} can be completed by setting a function of the data $\Y_t=\fb(\lambdab_{t}, \varepsilonb_t; \thetab_0)$ where $\{\varepsilonb_t\}_{t\in \Z}$ is an iid sequence. It is the case of the Vector Autoregressive Moving-Average models (VARMA), see \citet[Ch.~11]{lut2005}, or also the Multivariate Generalized Autoregressive Conditional Heteroskedasticity model (MGARCH); see \citet[Ch.~10]{fran2019}.  Although the MLE for model \eqref{mean equation} is sometimes readily computable when the joint conditional distribution function (cdf) can be specified (as in the case of VARMA and MGARCH models) this may not always be the case. For example, when count time series models are analyzed,  for models encompassed in eq.~\eqref{mean equation}, such as the multivariate Integer-valued GARCH (INGARCH) \citep{fok2020}, a Poisson distribution can be specified for the marginal probability mass functions (pmf) of the univariate process $Y_{i,t}$, for $i=1,\dots,d$. However, to obtain a complete joint cdf a particular joint pmf needs to be specified as well. The functional form of such a joint pmf can be unknown or too complex to be employed for the MLE.  
Indeed, the development of a multivariate count time series model would be based on the specification of a joint pmf distribution, so that the standard likelihood inference and testing procedures can be developed. 
The choice of a suitable multivariate version of the Poisson pmf is a challenging problem. Multivariate Poisson-type pmfs usually have a complicated closed form and the associated likelihood inference is theoretically and computationally cumbersome. Furthermore, in many cases,  the available multivariate Poisson-type pmfs   imply restricted models, which are of limited use in applications. 
 See \citet{inouye_2017} and \citet{fokianos_2021} for a  discussion on the choice of  multivariate count distributions and possible alternatives. Other approaches try to avoid the problem of specifying a multivariate Poisson distribution, like in \citet{fok2020}, where the joint distribution of the vector $\Y_t  $  is constructed by following a copula approach where all marginal distributions of $Y_{i,t}$ are univariate Poisson, conditionally to the past, with mean process $\lambda_{i,t}$, for $i=1,\dots,d$. However, this approach does not allow for performing a maximum likelihood estimation but requires appealing to the class of marginal QMLEs based on the linear exponential family that assumes independence between the univariate processes $Y_{i,t}$, for $i=1,\dots,d$. 
Analogous limitations appear when trying to model other non-negative series like binary and categorical time series. For example,  the Mixture
Transition Distribution (MTD) model of \cite{raftery1985model} requires the application of some form of marginal QMLE which assumes independence between observations. Similar limitations apply to multivariate logistic regressions. For further details the interested reader can see the related discussion in \cite{weiss2018}.

The marginal QMLE for multivariate time series can be generally written as the log-likelihood on marginal pdfs $Y_{i,t}|\mathcal{F}_{t-1}\sim q(\lambda_{i,t}(\thetab_0))$ coming from the exponential family of distributions \citep{wedderburn1974quasi} 

\begin{equation}
\hat \thetab_Q = \argmax_{\thetab \in \Theta } L^Q_T(\thetab)\,, \quad\quad	L^Q_T(\thetab)=T^{-1}\sum_{t=1}^{T} \sum_{i=1}^{d} \log q(\lambda_{i,t}(\thetab))\,.
\label{qmle}
\end{equation}
If the conditional mean vector $\lambdab_{t}(\thetab_0)$ is correctly specified then \eqref{qmle} is consistent and asymptotically normal for the unknown parameters $\thetab_0$ \citep{gourieroux1984pseudo}. Moreover, the QML estimator $\hat{\thetab}_Q$ can also be seen as the vector which solves the first-order conditions with the form

\begin{equation}
\Sb^Q_T(\thetab) = T^{-1} \sum_{t=1}^{T}\frac{\partial\lambdab_{t}(\thetab)'}{\partial\thetab}\mathbf{D}^*_t( \thetab)^{-1}\Big(\Y_t-\lambdab_{t}(\thetab)\Big) = \mathbf{0}\,,
\label{score qmle}
\end{equation}
where $\mathbf{S}^Q_T(\thetab)$ is the score of the QMLE, $\mathbf{D}^*_t(\thetab)$ is the $d\times d$ diagonal matrix with diagonal elements equal to $\nu^*_{i,t}(\thetab)$ for $i=1,\dots, d$ and $\nu^*_{i,t}(\thetab)$ is the single element of  the conditional pseudo-variance vector of the process
\begin{equation}
\nub^*(\Y_{t-1}, \Y_{t-2}, \dots; \thetab)=\nub^*_t(\thetab)\,, \quad t\in \Z
\label{var equation}
\end{equation}
with $ \nub^* : \R^\infty \times \Theta \to \left( 0, +\infty \right)^d $  being a measurable function. Equation~\eqref{var equation} is referred to as a pseudo-variance vector as it is not necessarily correctly specified.
From the property of the linear exponential family the pseudo-variance model depends on the mean model, i.e. $\nu^*_{i,t}(\thetab)=g(\lambda_{i,t}(\thetab))$ where $g(\cdot)$ is some measurable function. The possible value of $\nu^*_{i,t}$ is however restricted by the fact that it must match the conditional variance of the selected distribution in the exponential family. For example, in the QMLE employed in \citet{fok2020} all marginal distributions of $Y_{i,t}$ are univariate Poisson, conditionally to the past, so  $\nu^*_{i,t}(\thetab)=\lambda_{i,t}(\thetab)$, for $i=1,\dots,d$. In practice, one can easily
conceive that the true conditional variance may have other forms. The choice of an inadequate marginal pmf, or probability density function (pdf), may affect the efficiency of the QMLEs. Moreover, it is clear that the score of the QMLE described in \eqref{score qmle} does not directly include any joint dependence between the univariate processes because it is computed upon a quasi-likelihood which assumes contemporaneous independence among $Y_{i,t}$, for $i=1,\dots,d$. Therefore, the QMLE suffers from two main sources of inefficiency: constrained variance specification and exclusion of joint dependence parameters. The aim is twofold: improving the efficiency of QMLEs by reducing the impact of these two sources of inefficiency by  retaining the same number of assumptions on the model i.e. a semi-parametric approach assuming only the conditional mean  to be correctly specified.

\section{Multivariate Weighted Least Squares Estimators} \label{MWLE}

In order to mitigate the inefficiency of QMLEs a novel class of two-stage Multivariate Weighted Least Square Estimators (MWLSEs) of the mean parameters $\thetab_0$ is proposed. Given a matrix weight function $\W_t = \W(\Y_{t-1}, \Y_{t-2}, \dots)$, where $\W$ is a measurable function from $\R^\infty \to \R^{d \times d}$, a first-stage MWLSE is defined by 

\begin{equation}
\hat \thetab_1 = \argmin_{\thetab \in \Theta } L_T(\thetab, \W)
\label{first-stage estimator}
\end{equation}
where
\begin{equation}
L_T(\thetab, \W) =  T^{-1} \sum_{t=1}^{T}	l_t(\thetab, \W_t)\,, \quad \quad  l_t(\thetab, \W_t) = \Big(\Y_t-\lambdab_{t}(\thetab)\Big)^\prime \W_t\Big(\Y_t-\lambdab_{t}(\thetab)\Big)\,.
\label{likelihood 1 stage}
\end{equation}

For example, when $\W_t=\I$ where $\I$ is the identity matrix of suitable dimension, \eqref{likelihood 1 stage} corresponds to the multivariate version of the conditional Least Squares Estimator (LSE)  of \citet{klimko_1978conditional}. 
 It is well-known that the optimal choice of the weights $\W_t$ in weighted least squares is the inverse of the true conditional covariance matrix of the process, say $(\Vb_t^{-1})_{t \geq 1}$. In practice, $\Vb_t$ is generally unknown. Therefore, a conditional pseudo-covariance matrix is specified with a parametric specification of the form $\Vb^*_t (\taub_0) = \Vb^*(\Y_{t-1}, \Y_{t-2}, \dots; \taub_0)$, where $\Vb^*$ is a measurable function defined in $\R^\infty \to \R^{d \times d}$, which is not necessarily the true form of the covariance matrix. It depends on a pseudo-true parameter vector $\taub_0$ containing the true mean parameter vector $\thetab_0$ 
 and possibly some extra parameters $\gammab_0$. Define $\hat{\taub}=(\hat \thetab_1^\prime, \hat \gammab^\prime)^\prime $ the estimator of $\taub_0$ where $\hat \thetab_1$ is the first-stage estimator of $\thetab_0$ and $\hat \gammab$ is some previously obtained estimator of $\gammab_0$. Then, the working sequence of weights is estimated as $
\hat{\W}_t = \Vb^*_t( \hat{\taub})^{-1}$ and the two-stage MWLSE for the unknown mean parameters $\thetab_0$ is given by 

\begin{equation}
\hat \thetab = \argmin_{\thetab \in \Theta } L_T(\thetab, \hat{\W})\,, \quad \quad L_T(\thetab, \hat{\W}) =  T^{-1} \sum_{t=1}^{T}	l_t(\thetab, \hat{\W}_t)\,.
\label{second stage mwlse}
\end{equation}
Hence, the two-stage MWLSE defined in \eqref{second stage mwlse} is the solution of the first-order condition system

\begin{equation}
\Sb_T(\thetab, \hat \W) = T^{-1} \sum_{t=1}^{T}\frac{\partial\lambdab_{t}(\thetab)'}{\partial\thetab}\Vb_t^*( \hat{\taub})^{-1}\Big(\Y_t-\lambdab_{t}(\thetab)\Big)=\boldsymbol{0}\,.
\label{score mwlse}
\end{equation}
The first-order conditions \eqref{score qmle} and  \eqref{score mwlse} allow for a direct and intuitive comparison between QMLEs and MWLSEs. The first main difference is that in \eqref{score mwlse} there is no particular constraint on the shape of the univariate conditional pseudo-variances, $\nu^*_{i,t}(\taub)$, which are the elements in the main diagonal of $\Vb^*_t(\taub)$ for $\taub=(\thetab',\gammab')' \in \Tb$, where $ \Tb = \Theta \times \Gamma$ and $\Gamma$ is the parameter space of the extra parameters $\gammab$, whereas in \eqref{score qmle} the form of the pseudo-variances are constrained by the univariate conditional distribution specified in the quasi-likelihood. Moreover, the MWLSE allows the function $\Vb^*_t(\cdot)$ to depend on extra parameters apart from the mean vector parameters $\thetab$. Consequently, the MWLSE adds more flexibility in the choice of the variance form and this can lead to an improved efficiency estimator. A second difference lies in the possibility of including correlation effects in the estimation system \eqref{score mwlse} whereas the QMLE simply implies no correlation in the score \eqref{score qmle}, i.e. $\Vb^*_t(\taub)=\mathbf{D}^*_t( \thetab)$ is a diagonal working covariance matrix. For these additional features the MWLSE is expected to explain a larger part of the variability connected to the process of study compared to QMLE. 

The following theoretical results demonstrate that under correct specification of the conditional covariance matrix the two-stage MWLSE is asymptotically more efficient than the QMLE. This result will be further confirmed in Section~\ref{SEC simulations} with a simulated experiment. When a moderate or high degree of contemporaneous dependence is indeed present among the variables $Y_{i,t}$ for $i=1,\dots,d$ the two-stage MWLSE entails a relevant gain in efficiency with respect to the QMLE.

The MWLSEs require only the correct specification of the first conditional moment of the process instead of entirely specifying its conditional distribution. In this sense, they are semi-parametric estimators because, except for the first moment, they are totally agnostic about the distribution of the observations.  In this way the estimation is robust against possible misspecification of the conditional distribution and simplifies the inference when the full MLE is infeasible or too complex. Finally, MWLSEs present the extra advantage of being quite immediate to estimate since they require the call of two least squares estimations.

\begin{ex} (Example of working covariance specification) \rm
A possible specification of the working sequence of weights is
\begin{equation}
	\hat{\W}_t = \Vb^*_t( \hat{\taub})^{-1}=\mathbf{D}^*_t(\hat{\taub})^{-1/2}\mathbf{P}^*_t(\hat{\taub})^{-1}\mathbf{D}^*_t(\hat{\taub})^{-1/2}
	\label{optimal weights}
\end{equation}
where, analogously to Section~\ref{ineff QMLE}, $\mathbf{D}^*_t(\hat{\taub})$ is the $d\times d$ diagonal matrix with diagonal elements equal to $\nu^*_{i,t}(\hat{\taub})$ for $i=1,\dots, d$ and $\nu^*_{i,t}(\taub)$ is defined as 
\begin{equation}
	\nub^*(\Y_{t-1}, \Y_{t-2}, \dots; \taub)=\nub^*_t(\taub)\,, \quad t\in \Z\,.
	\label{var equation mwlse}
\end{equation}
Differently from \eqref{var equation} the pseudo-variance is not constrained to be some pre-specified function of the mean and can depend on extra parameters. Moreover, $\Pb^*_t(\taub)$ is a $d \times d$ matrix referred to as a pseudo-correlation matrix as it is not necessarily correctly specified. It is a working correlation matrix whose structure can be defined by the researcher and whose single element has a functional form of the type 
\begin{equation}
	r^*_{ij}(\Y_{t-1}, \Y_{t-2} \dots; \taub)= r^*_{ij,t}(\taub) \,, \quad \text{for} \quad  i,j=1,\dots,d
\end{equation}
where $r^*_{ij,t} :  \R^\infty \times \Tb \to (-1,1)$.  
 Conversely, the QMLE implies no correlation in the score \eqref{score qmle}, i.e. $\mathbf{P}^*_t( \taub)=\I$. 
 It is also possible to provide a simplified specification with constant correlation structure where $\mathbf{P}^*_t( \taub)=\mathbf{P}^*( \taub)$ for all $t\in\Z$ whose single element has the form $r^*_{ij}(\Y_{1}, \dots, \Y_{T}; \taub)=r^*_{ij}(\taub)$ with  $r^*_{ij} : \R^T \times \Tb \to (-1,1)$.
\end{ex}

\subsection{Asymptotic results for MWLSEs}

The following results establish the consistency and asymptotic normality of the first- and second-stage MWLSEs. Some additional notation is required for the asymptotic results of  MWLSEs. For a vector
function $\fb : \Theta \to \R^d$ and a matrix
function $\ffb : \Theta \to \R^{d\times d}$, the supremum norms are defined as $\lnorm{\fb}_\Theta = \sup_{\thetab \in \Theta} \lnorm{\fb(\thetab)}$ and $\vertiii{\ffb}_\Theta = \sup_{\thetab \in \Theta} \vertiii{\ffb(\thetab)}$ where $\lnorm{\cdot}$ is the Euclidean norm and $\vertiii{\cdot}$ is the spectral norm. In general $\lnorm{\cdot}_p$ defines the $l^p$ norm. 
 For any symmetric matrix $\X$, $\X > 0$  $(\X\geq 0)$ is a shorthand for ``$\X$ is a positive (semi-)definite matrix". 
 The following assumptions are required to establish asymptotic results for $\hat \thetab_1$, the first-stage MWLSE.

\begin{enumerate}[label=\textbf{A\arabic*}]
\item The process $\{ \Y_t, t \in \Z \}$  is strictly stationary and ergodic. %
\label{Ass. stationarity}
\item \label{Ass. uniform moment} 
$\lambdab_t(\cdot)$ is a.s. continuous and the set $\Theta$ is compact. Moreover, $\E \lnorm{l_t(\thetab, \W_t)}_\Theta    < \infty$.
\item $\W_t > 0$ for all $t \in \Z$. 
  \label{Ass. pos def.}
\item \label{Ass. identification} $\lambdab_t(\thetab) = \lambdab_t(\thetab_0)$ a.s. if and only if $\thetab=\thetab_0$. Moreover, $\lambdab_t(\thetab_0)$ belongs a.s. to the interior of $\Lambdab$.

\item \label{Ass. moment second derivative}
$\lambdab_t(\cdot)$ has a.s. continuous second-order derivatives and for $k,l=1,\dots,m$ 

$$ \E \lnorm{ \frac{\partial \lambdab_t' }{\partial \theta_k} \W_t  \frac{\partial \lambdab_t }{\partial \theta_l}}_\Theta < \infty\,, ~~~~ \E \lnorm{ \frac{\partial^2 \lambdab_t'}{\partial \theta_k \partial \theta_l} \W_t  \left( \Y_t - \lambdab_t \right) }_\Theta < \infty  \,.$$

\item \label{Ass. hessians} The following matrices exist $$\G(\thetab_0, \W) = \E\left[  \frac{\partial\lambdab_{t}(\thetab_0)'}{\partial\thetab} \W_t \Vb_t \W_t \frac{\partial\lambdab_{t}(\thetab_0)}{\partial\thetab} \right] \,, \quad \Hb(\thetab_0, \W)=\E \left[ \frac{\partial\lambdab_{t}(\thetab_0)'}{\partial\thetab} \W_t \frac{\partial\lambdab_{t}(\thetab_0)}{\partial\thetab} \right] $$ with  $\Hb(\thetab_0, \W) > 0$.

\item $\thetab_0 \in \dot{\Theta}$, where $\dot{\Theta}$ is the interior of $\Theta$. \label{Ass. interior}

%
%
%
%
\end{enumerate}
The described assumptions are quite standard in time series and imply that the conditional mean process is identifiable, stable, differentiable, and has existing and invertible limiting matrices. See Section~\ref{SEC examples} for some examples with specific models for the conditional mean.

\begin{theorem} \label{Thm. first stage}
Consider the first-stage MWLSE \eqref{first-stage estimator}. Under conditions~\ref{Ass. stationarity}-\ref{Ass. identification}

\begin{equation} \label{consistency}
	\hat \thetab_1 \xrightarrow{} \thetab_0\,, \quad a.s. \quad T \to \infty\,.
\end{equation}
Moreover, if also \ref{Ass. moment second derivative}-\ref{Ass. interior} hold, as $T \to \infty$
\begin{equation} \label{asymp. normality}
	\sqrt{T} \left( \hat \thetab_1 - \thetab_0 \right)  \xrightarrow{d} N\left( 0, \Sigmab\right) \,, \quad  \Sigmab=\Sigmab(\thetab_0, \W)=\Hb^{-1}(\thetab_0, \W) \G(\thetab_0, \W) \Hb^{-1}(\thetab_0, \W)\,.
\end{equation}
\end{theorem}

Let $\W_t = \Vb^*_t(\taub_0)^{-1}$ so that the working weights for the two-stage MWLSE will be $\hat \W_t=\Vb^*_t(\hat \taub)^{-1}$
Note that asymptotic results of the two-stage MWLSE cannot be directly deduced from those of the first-stage MWLSE because, contrary to $\W_t$, $\hat \W_t$ is not $\Fb_{t-1}$-measurable since it depends on parameters $\hat \taub$ estimated using the whole sample of data.
In order to establish analogous results for the two-stage MWLSE some additional notation and assumptions are required. Recall that $\taub=(\thetab^\prime, \gammab^\prime)^\prime$ where $\taub \in \Tb$. 
Define by $\vertiii{\cdot}_F$ the Frobenius norm 
and $\kb_t(\taub) = \text{vec}(\Vb^*_t(\taub)^{-1})$. 

\begin{enumerate}[label=\textbf{B\arabic*}]

\item \label{Ass. weights}  $\kb_t(\cdot)$ is a.s. continuously differentiable and the set $\Tb$ is compact. In addition,
$$\E\, \vertiii{\frac{\partial \kb_t}{\partial \taub}}_\Tb \, \lnorm{\Y_t -\lambdab_t }_\Theta^2 < \infty\,.$$

\item \label{Ass. moments consistency second stage}  $\hat \gammab \to \gammab_0$ a.s. as $T \to \infty$.

\item \label{Ass. hessian second stage}  For $k,l=1,\dots,m$

$$\E\,\vertiii{\frac{\partial \kb_t}{\partial \taub}}_\Tb \, \vertiii{\frac{\partial \lambdab_t }{\partial \theta_k}   \frac{\partial \lambdab_t' }{\partial \theta_l}}_{F,\Theta} < \infty\,, ~~~~ \E\,\vertiii{\frac{\partial \kb_t}{\partial \taub}}_\Tb \, \vertiii{ \left( \Y_t -\lambdab_t \right)  \frac{\partial^2 \lambdab_t' }{\partial \theta_k \partial \theta_l}}_{F,\Theta} < \infty\,, $$
$$\E\,\vertiii{\frac{\partial \kb_t}{\partial \taub}}_\Tb \, \vertiii{\left( \Y_t - \lambdab_t(\thetab_0) \right) \frac{\partial \lambdab_t(\thetab_0)'}{\partial \theta_k}  }_F< \infty\,.$$

\item \label{Ass. bounded in prob} $\sqrt{T}(\hat \taub - \taub_0)=O_p(1)$.


\end{enumerate}
The additional assumptions are quite common in two-stage estimators. These require differentiability of the weighting sequence with respect to its static parameters and that the previous stage estimators should be reasonably well-behaved. 
Moreover when $\W_t$ is well specified, i.e. $\W_t=\Vb_t^{-1}$, Assumption~\ref{Ass. hessians} simplifies as follows.
\begin{enumerate}[label=\textbf{A6*}]
\item \label{Ass. hessians optimal} The matrix  $$\G = \E \left[ \frac{\partial\lambdab_{t}(\thetab_0)'}{\partial\thetab} \Vb_t^{-1} \frac{\partial\lambdab_{t}(\thetab_0)}{\partial\thetab} \right] $$ exists with  $\G > 0$.
\end{enumerate}

\begin{theorem} \label{Thm. second stage}
Consider the two-stage MWLSE \eqref{second stage mwlse}. Under conditions~\ref{Ass. stationarity}-\ref{Ass. identification} and \ref{Ass. weights}-\ref{Ass. moments consistency second stage}
\begin{equation} \label{consistency second stage}
	\hat \thetab \xrightarrow{} \thetab_0\,, \quad a.s. \quad T \to \infty\,.
\end{equation}
Moreover, if also \ref{Ass. moment second derivative}-\ref{Ass. interior} and \ref{Ass. hessian second stage}-\ref{Ass. bounded in prob} hold, as $T \to \infty$
\begin{equation} \label{asymp. normality second stage}
	\sqrt{T} \left( \hat \thetab - \thetab_0 \right)  \xrightarrow{d} N\left( 0, \Sigmab \right)\,, \quad  \Sigmab = \Sigmab(\thetab_0, \W)=\Hb^{-1}(\thetab_0, \W) \G(\thetab_0, \W) \Hb^{-1}(\thetab_0, \W) \,.
\end{equation}
If in addition the conditional variance matrix is correctly specified, 
that is $\Vb^*_t(\taub_0)=\Vb_t$, then \ref{Ass. hessians} can be replaced by \ref{Ass. hessians optimal} and 
\begin{equation} \label{asymp. normality second stage eff}
	\sqrt{T} \left( \hat \thetab - \thetab_0 \right)  \xrightarrow{d} N\left( 0, \G^{-1}\right) \,,
\end{equation}
where $\Sigmab - \G^{-1} \geq 0 $.

\end{theorem}
The previous result shows that in the special case in which the conditional variance matrix is well-specified then the two-stage MWLS estimator $\hat \thetab$ gains in efficiency, in the sense that the matrix $\Sigmab - \G^{-1} $ is positive semi-definite. This also implies that the MWLSE is (asymptotically) more efficient than the QMLE. For example, consider the Poisson QMLE, $\hat \thetab_P=\argmax_{\thetab \in \Theta } L^P_T(\thetab)$ where $L^P_T(\thetab)$ is defined as in \eqref{qmle} with $q(\cdot)$ being the Poisson pmf. The same arguments of the proof of Theorem~\ref{Thm. first stage} and \citet[eq.~34]{fra2016} establish the consistency and asymptotic normality of the Poisson QMLE.
\begin{equation*} 
	\sqrt{T} \left( \hat \thetab_P - \thetab_0 \right)  \xrightarrow{d} N\left( 0, \Sigmab_P \right)\,, \quad  \Sigmab_P =\Hb_P^{-1} \G_P\Hb_P^{-1} \,,
\end{equation*}
with
$$\G_P = \E\left[  \frac{\partial\lambdab_{t}(\thetab_0)'}{\partial\thetab} \mathbf{D^*}_t^{-1}( \thetab_0) \Vb_t \mathbf{D^*}_t^{-1}( \thetab_0) \frac{\partial\lambdab_{t}(\thetab_0)}{\partial\thetab} \right] \,, \quad \Hb_P=\E \left[ \frac{\partial\lambdab_{t}(\thetab_0)'}{\partial\thetab} \mathbf{D^*}_t^{-1}( \thetab_0) \frac{\partial\lambdab_{t}(\thetab_0)}{\partial\thetab} \right]\,, $$
where $\mathbf{D^*}_t( \thetab_0)$ is a diagonal matrix with elements $\lambda_{i,t}(\thetab_0)$ for $i=1,\dots,d$. 
Analogous arguments and \citet[eq.~A.8]{aknouche2021count} yield similar asymptotic results for the exponential QMLE, which can be defined as $\hat \thetab_E=\argmax_{\thetab \in \Theta } L^E_T(\thetab)$ where $L^E_T(\thetab)$ is as in \eqref{qmle} with $q(\cdot)$ being the exponential pdf.
\begin{equation*} 
	\sqrt{T} \left( \hat \thetab_E - \thetab_0 \right)  \xrightarrow{d} N\left( 0, \Sigmab_E \right)\,, \quad  \Sigmab_E =\Hb_E^{-1} \G_E\Hb_E^{-1} \,,
\end{equation*}
with
$$\G_E = \E\left[  \frac{\partial\lambdab_{t}(\thetab_0)'}{\partial\thetab} \mathbf{D^*}_t^{-2}( \thetab_0) \Vb_t \mathbf{D^*}_t^{-2}( \thetab_0) \frac{\partial\lambdab_{t}(\thetab_0)}{\partial\thetab} \right] \,, \quad \Hb_E=\E \left[ \frac{\partial\lambdab_{t}(\thetab_0)'}{\partial\thetab} \mathbf{D^*}_t^{-2}( \thetab_0) \frac{\partial\lambdab_{t}(\thetab_0)}{\partial\thetab} \right]\,. $$
Note that $\G_P=\G(\thetab_0, \W)$, $\Hb_P=\Hb(\thetab_0, \W)$ with $\W_t=\mathbf{D^*}_t^{-1}( \thetab_0)$ and $\G_E=\G(\thetab_0, \W)$, $\Hb_E=\Hb(\thetab_0, \W)$ with $\W_t=\mathbf{D^*}_t^{-2}( \thetab_0)$.
Therefore, $\Sigmab_P$ and $\Sigmab_E$ are special cases of $\Sigmab$. This entails the following result.

\begin{corollary}
Assume $\Y_t \geq \zerob$ and that the WLSEs and QMLEs are consistent and asymptotically normal. If the conditional variance matrix is well specified, the two-stage MWLSE is asymptotically more efficient than the QMLEs, in the sense that the matrices $\Sigmab_P - \G^{-1}$ and  $\Sigmab_E - \G^{-1}$ are positive semi-definite.
\end{corollary}

Another advantage of MWLSEs is that they avoid boundary problems that may instead appear when QMLEs are employed. 
For example, when modelling count or positive time series by  Poisson or exponential QMLE, the expression of the quasi-likelihood requires the computation of $\log(\lambdab_t(\thetab))$ for all $\thetab \in \Theta$, otherwise $\hat \thetab_P$ and $\hat \thetab_E$ are not well-defined. Moreover, following the arguments of \citet[Sec.~2.2.2]{fra2016} the asymptotic Gaussian distribution of the QMLE may break if the dynamics of $\lambdab_t(\thetab)$ is linear and some of the values in $\thetab_0$ are zero (because $\thetab_0 \notin \dot{\Theta}$).
Instead, when using MWLSEs it is possible to have $\lambdab_t(\thetab)<0$ for some $\thetab$, although one must have $\lambdab_t(\thetab_0) \geq 0$ for non-negative observations. This result implies that  $\thetab_0$  may include one or more null components while still belonging to the interior of the parameter space $ \dot{\Theta}$.  Therefore, boundary problems for $\thetab_0$ are avoided. See \citet[Rem.~2.1]{francq_2021two} for more details. 

It is now shown that the MWLSE is asymptotically efficient when the true distribution of $\Y_t$ belongs to the wide class of the linear multivariate exponential distributions. With respect to some $\sigma$-finite measure $\mu$ (in general the Lebesgue measure or the counting measure), let $f_{\lambdab}$ be the density of a real random vector $\Y$ of mean 
\[
\lambdab = \int_\Yc \yb f_{\lambdab}(\yb) \, d\mu(\yb)
\]
where $\Yc \subseteq \R^d$ is the domain of $\Y$.
Let $\Lambdab$ be a non-empty open subspace of $\R^d$. Then the set $\{f_{\lambdab}, \lambdab \in \Lambdab\}$ constitutes a $d$-parameter linear multivariate exponential family if for all $\lambdab \in \Lambdab$
\begin{equation} \label{ef}
	f_{\lambdab}(\yb) = h(\yb) e^{\etab(\lambdab)'\yb - a(\lambdab)}
\end{equation}
with $h: \R^d \to \R$ and some twice differentiable maps $\etab: \R^d \to \R^d$ and $a: \R^d \to \R$.

\begin{theorem} \label{Thm. efficiency ef}
Assume \ref{Ass. stationarity}-\ref{Ass. uniform moment}, \ref{Ass. identification}-\ref{Ass. moment second derivative} and \ref{Ass. interior} hold. Suppose that the conditional distribution of $\Y_t$ given $\lambdab_t = \lambdab$ has the linear exponential form \eqref{ef}. 
Assume also that the columns of the matrices $\partial\etab(\lambdab)/\partial\lambdab$ and $\partial\lambdab_{t}(\thetab_0)/\partial\thetab$  are linearly independent. Then, if the conditional variance matrix is well specified, the two-stage MWLSE is asymptotically as efficient as the MLE of $\thetab_0$.
\end{theorem}

\section{Conditional mean modeling examples}

\label{SEC examples}

In this section the established asymptotic results for the first- and second-stage MWLSEs are applied to some specific models of interest under low-level conditions.

One of the possible alternatives for modelling multivariate integer-valued time series, such as counts in epidemiology,  is the multivariate INGARCH model \citep{fok2020}
\begin{equation}
	\Y_t=\mathbf{N}_t(\lambdab_t), ~~~ \lambdab_t=\mathbf{c}+\A\lambdab_{t-1}+\B\Y_{t-1}\,,
	\label{ingarch}
\end{equation} 
where $\mathbf{c}$ is a vectorial intercept of positive unknown parameters and $\A$ and $\B$ are $d \times d$ unknown parameter matrices assumed to be non-negative such that $\lambda_{i,t} > 0$ for all $i$ and $t$. $\mathbf{N}_t(\lambdab_t)$ is a sequence of $d$-dimensional iid marginally Poisson count processes, with intensity 1, counting the number of events in the interval of time $[0,\lambda_{1,t}]\times\dots\times[0,\lambda_{d,t}]$ and whose structure of dependence is modelled through a copula construction $C(\dots; \rho)$ on their associated exponential waiting times random variables, and which depends on one unknown parameter, say $\rho$, capturing the contemporaneous correlation among the variables. The complete structure of the data generating process (dgp) is described in \citet[p.~474]{fok2020} and \cite{armillotta_fokianos_2022_testing,armillotta_fokianos_2021}. Since the copula construction in the dgp is imposed on unknown latent variables (exponential waiting times) the joint conditional pmf resulting from \eqref{ingarch} does not have a closed form. This rules out the possibility of computing a full MLE. However, the univariate conditional pmfs $q(\cdot)$ are Poisson, i.e. $Y_{i,t}|\mathcal{F}_{t-1}\sim Pois(\lambda_{i,t}(\thetab_0))$, for $i=1,\dots,d$.

For a matrix $\X \in \R^{d \times d}$ define by $\lambda_{\max}(\X)$ ($\lambda_{\min}(\X)$) the maximum (minimum) eigenvalue of $\X$, by $\rho(\X)$ the spectral radius of $\X$, and by $\X \succeq 0$ that  $\X$ as non-negative elements. For a vector $x \in \R^d$, the notation $x  \nsim 0$ denotes that all elements of $x$ are not 0. Define $1_d$ a $d$-dimensional vector of ones. Consider the following assumption for the first-stage weighting matrix.
\begin{enumerate}[label=\textbf{C\arabic*}]
	\item 	\label{pos def bound} $\lambda_{\min}(\W_t^{-1})  \geq \underline{\lambda} > 0$.
\end{enumerate} 

\begin{proposition} \label{Prop. mult ingarch}
	Assume that $\Y_t$ follows model \eqref{ingarch} where $\lambdab_t=\lambdab_t(\thetab_0)$. Let
 $\thetab_0 \in \dot{\Theta}$ where $\Theta$ is a compact parameter set such that $ \mathbf{c} \succ 0$, $\A \succeq 0$, $\B \succeq 0$ for any $\thetab \in \Theta$, $|\B_0|_e 1_d \nsim 0$, and $\rho(\A_0 + \B_0) < 1$. Then, for any sequence of weights $(\W_t)$ satisfying \ref{pos def bound}, Theorem~\ref{Thm. first stage} holds. 
\end{proposition}
The spectral radius condition implies stationarity and ergodicity of the count process. it is shown that the condition $|\B_0|_e 1_d \nsim 0$ is required for the identifiability of the feedback parameters $\A$. Finally, \ref{pos def bound} guarantees positive definiteness of the weighting sequence with bounded spectral norm.

For the second-stage estimator, let us take the weighting matrix $\hat \W_t$ with structure \eqref{optimal weights} where $\nub^*_t(\hat \taub)=\lambdab_t(\hat \thetab_1)$ (by the Poisson property) and $\mathbf{P}^*_t(\hat{\taub})=\Pb^*(\hat \thetab_1)= (1-r^*(\hat \thetab_1))\I_d+r^*(\hat \thetab_1) \mathbf{J}_d$, where $\I_d$ and $\mathbf{J}_d$ are respectively an identity matrix and a matrix of ones of dimensions  $d\times d$, whereas $r^*(\hat \thetab_1)=2\sum_{i=1}^{d}\sum_{j>i} r^*_{ij}(\hat{\thetab}_1)/[d(d-1)]$ and for all $i,j=1,\dots, d$ one has that $r^*_{ij}(\hat \thetab_1)=T^{-1} \sum_{t=1}^{T}u^*_{i,t}(\hat \thetab_1)u^*_{j,t}(\hat \thetab_1)$ with  $u^*_{i,t}(\hat \thetab_1)=\left[ Y_{i,t}-\lambda_{i,t}(\hat \thetab_1)\right]/\lambda_{i,t}(\hat \thetab_1)^{1/2}$ being the Pearson residuals. Such a working correlation matrix is appealing because $r^*(\hat \thetab_1)$ can be precisely estimated using the whole history of all the pairs $(i,j)$. Moreover, the equicorrelation structure has an analytical form of the inverse which allows the practitioner to further speed-up numerical evaluations. Indeed, sometimes the inversion of the $d \times d$ matrix $\Pb^*(\cdot)$ may be unfeasible for numerical reasons. Furthermore, since only the inverse of such a matrix would be required by the proposed estimating procedure, it is suggested to choose a working correlation matrix structure where an analytical form for the inverse is  known.   
Assumption \ref{pos def bound} can be substituted by a low-level condition. 

\begin{enumerate}[label=\textbf{C1*}]
	\item 	\label{pos def bound equi} $ r^*(\thetab) \in [\underline{r},\bar r]$ for any $\thetab \in \Theta$, where $ \underline{r} > -(d-1)^{-1} $ and $\bar r <1$.
\end{enumerate} 

\begin{proposition} \label{Prop. mult ingarch second stage}
	Under the assumptions of Proposition~\ref{Prop. mult ingarch}, where \ref{pos def bound} is substituted by \ref{pos def bound equi}, Theorem~\ref{Thm. second stage} holds. 
\end{proposition}

The bounds in condition \ref{pos def bound equi} ensure positive definiteness of the working (equi)-correlation matrix and hence satisfy \ref{pos def bound} for the second-stage weighting matrix. 

More complex univariate count distributions can be accommodated in this framework, like the negative binomial distribution, by applying a modification of the data generating algorithm, as described in \cite{guo_zhu_2024}; see also \cite{guo_zhu_soft_2024}. In this case the distributions depend on extra dispersion parameters $\gamma_i$ which can be estimated at the first stage by employing the moment estimator proposed in \cite{gourieroux1984pseudo}. See also \citet[eq.~3.9]{francq_2021two}.



For modeling multivariate positive time series, such as vectors of durations or volumes, a multivariate ACD model  \citep{hautsch2012econometrics} can be employed
\begin{equation}
	\Y_t=\lambdab_t \odot \Zb_t, ~~~ \lambdab_t=\mathbf{c}+\A\lambdab_{t-1}+\B\Y_{t-1}\,,
	\label{acd}
\end{equation} 
where $\odot$ denotes the Hadamard product and $Z_{i,t} \sim  Exp(1)$ are iid over time, implying that $Y_{i,t}$ is exponentially distributed (conditionally on the past) with mean $\lambda_{i,t}$, for $i=1,\dots,d$, and whose joint structure of dependence can modelled following \cite{debaly_truquet_2023} through a copula construction $C(\dots; \rho)$ on the exponential margins, with a copula parameter $\rho$ capturing the contemporaneous correlation among the variables. Note that the same restrictions imposed in \eqref{ingarch} on the static parameters apply here since $\lambda_{i,t} > 0$ for all $i$ and $t$. The same structure of the weighting matrix described above for the two-stage estimator of the multivariate INGARCH model is employed here with $\nub^*_t(\hat \taub)=\lambdab_t^2(\hat \thetab_1)$ due to the exponential margins.

\begin{proposition} \label{Prop. mult acd}
	Assume that $\Y_t$ follows model \eqref{acd} where $\lambdab_t=\lambdab_t(\thetab_0)$ and $\E\lnorm{\Y_t}_4^4 <\infty$. Then, under 
	the assumptions of Propositions~\ref{Prop. mult ingarch}-\ref{Prop. mult ingarch second stage}, Theorems~\ref{Thm. first stage}-\ref{Thm. second stage} hold. 
\end{proposition}

\section{Numerical experiments}
\label{SEC simulations}

\begin{figure}[h]
\begin{center}
	\includegraphics[width=0.65\linewidth, height=0.45\textheight]{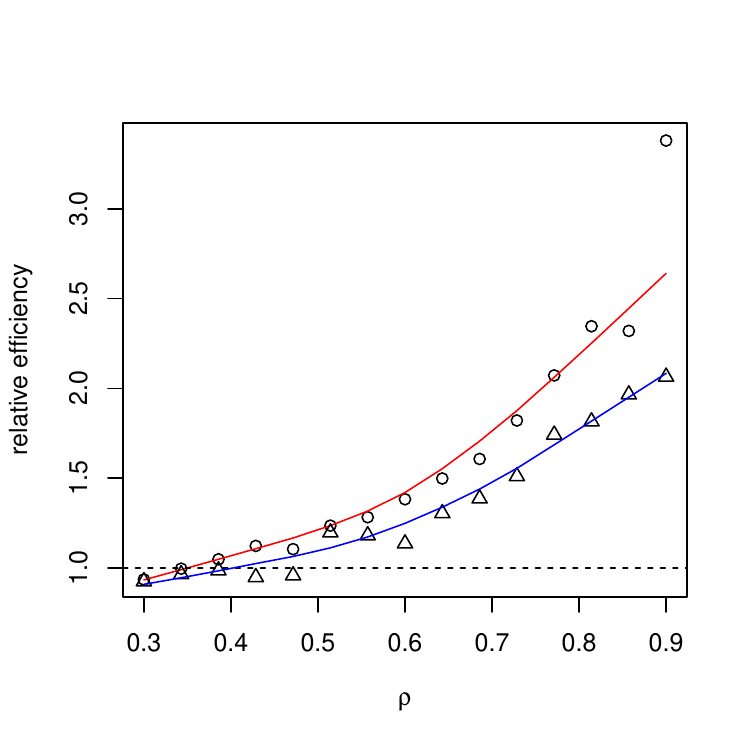}
	\caption{Plot for $e(\hat{\thetab}_{P},\hat{\thetab})$ versus values of copula parameter $\rho$, with working constant equicorrelation  matrix. Data generated by model \eqref{ingarch}
		with 500 simulations. Triangles: $d=2,T=100$. Points: $d=4,T=200$. Blue line: LOWESS smoother at $d=2,T=100$. Red line: LOWESS smoother at $d=4,T=200$. Dashed line: horizontal line $e(\hat{\thetab}_{P},\hat{\thetab})=1$.}%
	\label{rel_eff}
\end{center}
\end{figure}

In this section a numerical experiment using Monte Carlo simulations is presented to study the performance of QMLEs and MWLSEs. Data are generated from the count process described in \eqref{ingarch} with different dimension $d$ and sample size $T$, using the algorithm depicted in \citet[p.~474]{fok2020} and the Gaussian copula for $C(\dots; \rho)$; data are produced by employing a copula parameter $\rho$ selected by an equidistant grid of values in the interval $[0.3,0.9]$ using the \texttt{R} software \citep{armillotta2022r}. For the first-stage estimation let $\W_t=\I$ so that $\hat{\thetab}_1$ is the LSE. For the second-stage estimator the same structure described in Section~\ref{SEC examples} is retained. In this case no additional parameters are employed so $\hat{\taub}=\hat{\thetab}_1$.
 Then, the estimation performance of the Poisson QMLE versus the two-stage MWLSE are compared, by measuring their relative efficiency, with the relative Mean Square Error, $e(\hat{\thetab}_P,\hat{\thetab})=\sum_{s=1}^{S}||\hat{\thetab}_{P,s}-\thetab||^2/\sum_{s=1}^{S}||\hat{\thetab}_s-\thetab||^2$, where $S=500$ is the number of simulations performed and $\hat{\thetab}_s$ is the estimator associated with the replication $s$. Clearly, $e(\hat{\thetab}_P,\hat{\thetab})>1$ shows improved efficiency of $\hat{\thetab}$ when compared to $\hat{\thetab}_P$. The same comparison can be done marginally for each parameter of the model $e(\hat{\thetab}_{P,h},\hat{\thetab}_h)=\sum_{s=1}^{S}(\hat{\thetab}_{P,h,s}-\thetab_h)^2/\sum_{s=1}^{S}(\hat{\thetab}_{h,s}-\thetab_h)^2$, for $h=1,\dots,m$. The results of the Monte Carlo simulations are summarized in Figure~\ref{rel_eff}.
When the data are generated with a low correlation ($\rho < 0.4$) the two estimators show similar performance. However, when the data are generated with a moderate or strong correlation (copula parameter $\rho > 0.4$) the two-stage MWLSE is relatively more efficient than the QMLE. The improvement in efficiency is larger as the correlation becomes stronger and tends to grow even by increasing dimension and time size ($d,T$). This would be expected since the QMLE becomes a poor approximation of the true likelihood as the dependence structure of the multivariate count process $\Y_t$ becomes stronger, whereas the specified MWLSE methodology appears to be able to account for a significant part of the correlations among the counts, even though the working correlation does not reflect the true correlation structure of the data. Analogous results are obtained by comparing the marginal efficiencies $e(\hat{\thetab}_{Q,h},\hat{\thetab}_h)$, therefore they are omitted. 




\section{Modeling IT companies' stock return directions} \label{SEC: application}

\begin{figure}[h]
\begin{center}
	\includegraphics[width=0.7\linewidth]{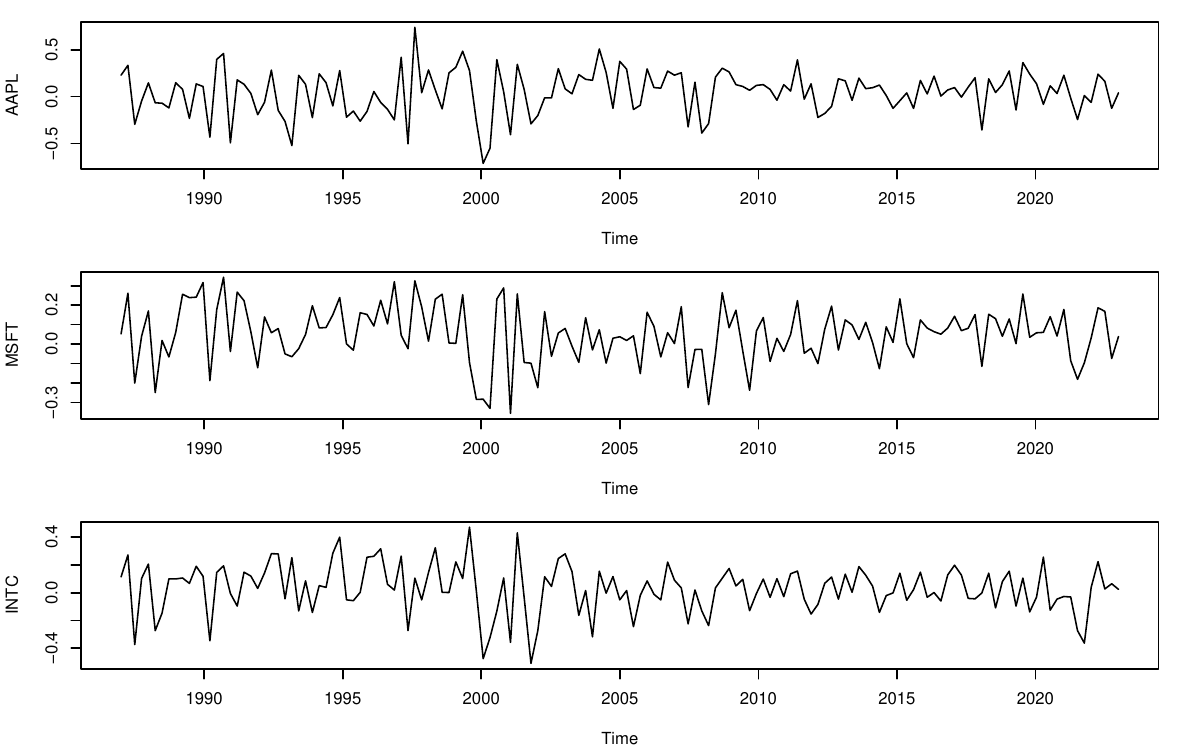}
	\caption{Plot of returns for three main IT companies: Apple (AAPL), Microsoft (MSFT) and Intel (INTC).}%
	\label{plot returns}
\end{center}
\end{figure}

Binary time series are of utmost importance in financial applications. Indeed, in financial time series analysis, it is well known that the expected values of the stock returns are quite difficult to predict, as they typically behave like white noise sequences. However, the volatility of the returns can generally be forecasted; this implies that also the sign of the returns can be predicted, see for instance
\cite{christoffersen2006financial} and \cite{christoffersen1direction}. \cite{breen1989economic}, among others, found that for excess stock returns only their direction is predictable. Moreover, in general it is possible to have sign dependence without conditional mean dependence. See also \cite{moysiadis2014binary} for additional references.
These arguments motivate the specification of a multivariate binary time series model for stock return signs. The returns are computed using the logarithmic formula $Re_{i,t} = \log(Pr_{i,t}) - \log(Pr_{i,t-1})$ where $Pr_{i,t}$ is the closing price for the stock of company $i$ at time $t$. Then, the multivariate binary time series is constructed by setting $Y_{i,t}=1$ if $Re_{i,t}\geq 0$, i.e. when the  return of the company $i$ at time $t$ is positive (equivalently the price change is positive) and $Y_{i,t}=0$ otherwise, i.e. when the price change is negative. From the general model \eqref{mean equation} the following Multivariate Binary Autoregression (MBAR) is proposed
\begin{equation}
\label{Bernoulli model}
\pb_t = \mathbf{c} +  \A \pb_{t-1} + \B \Y_{t-1}
\end{equation}
where $\Y_t \in \left\lbrace 0,1 \right\rbrace ^d$, $\pb_t = (p_{1,t},\dots,p_{i,t},\dots,p_{d,t})^\prime$ and $\
p_{i,t}=\P(Y_{i,t}=1|\Fb_{t-1})=\lambda_{i,t}$ is the conditional mean of binary data and therefore the probability of a positive price change. The unknown parameter vector and matrices should be constrained such that $ 0 < p_{i,t} < 1$ for all $i$ and $t$. For example, a simple sufficient condition is $\vertiii{\mathbf{C}+\A+\B}_{\infty}<1$ where $\mathbf{C}$ is a diagonal matrix hosting the elements of $\mathbf{c}$ and $\vertiii{\cdot}_\infty$ is the matrix infinity norm.
Model \eqref{Bernoulli model} is applied to the logarithmic returns of the quarterly closing prices of three main IT companies' stocks: Apple (AAPL), Microsoft (MSFT) and Intel (INTC), simply codified by $i=1,2,3$, respectively, for the period from Q1-1987 to Q3-2023. 
The returns data are plotted in Figure~\ref{plot returns} and can be freely downloaded from  \url{http://finance.yahoo.com}. The returns show several fluctuations in price change directions. In particular, negative sign drops can be appreciated around 2000, corresponding to the dot-com bubble, in 2008 with the rise of the Great crisis and after the COVID-19 pandemic due to inflation and recent Ukrainian war instabilities.

\cite{christoffersen2006financial} showed that sign dependence
is not likely to be found in very high-frequency (e.g. daily/weekly) or very low-frequency (e.g. annual) returns; instead, it is more likely to be found at intermediate return horizons. Therefore, stock prices are aggregated at the quarterly level. This leaves the sample with dimension $n=147$.
Following standard practice for count time series \citep{Heinen(2003)}, \cite{Leeetal(2017)} define $\A$ as a diagonal matrix for parsimony since the time series is short. 
Model~\eqref{Bernoulli model} is estimated by employing the proposed two-stage MWLSE with working weight \eqref{optimal weights}, LSE as first-stage estimator, pseudo-variance specification $\nu^*_{i,t} = p_{i,t}(1-p_{i,t})$ and constant equicorrelation which is estimated to be $ r^*(\hat \thetab_1) \approx 0.32$.  The result of the estimation is summarized in \eqref{BerWLS estimates} and is compared with estimation performed by QMLE \eqref{qmle} with Bernoulli marginals in \eqref{BerQML estimates}. The symbols $^\dagger$, $^*$ and $^{**}$ refer to coefficients significantly different from 0 at 0.1\%, 0.05\% and 0.01\% confidence level, respectively.

\begin{equation}
\label{BerQML estimates}
\begin{bmatrix}
	\hat p^B_{1,t}  \\
	\hat p^B_{2,t}   \\
	\hat p^B_{3,t}
\end{bmatrix} 
=  \begin{bmatrix}
	0.567  \\
	0.532   \\
	0.352 
\end{bmatrix}
+  \begin{bmatrix}
	0.000 & - & - \\
	- & 0.146 & - \\
	- & - & 0.393^\dagger
\end{bmatrix} 
\begin{bmatrix}
	\hat p^B_{1,t-1}  \\
	\hat p^B_{2,t-1}   \\
	\hat p^B_{3,t-1}
\end{bmatrix} + 
\begin{bmatrix}
	0.020 & 0.022 & 0.048\\
	0.000 & 0.077^* & 0.000 \\
	0.000 & 0.000 & 0.022
\end{bmatrix}
\begin{bmatrix}
	Y_{1,t-1}  \\
	Y_{2,t-1}   \\
	Y_{3,t-1}
\end{bmatrix}
\end{equation}

\begin{equation}
\label{BerWLS estimates}
\begin{bmatrix}
	\hat p_{1,t}  \\
	\hat p_{2,t}   \\
	\hat p_{3,t}
\end{bmatrix} 
=  \begin{bmatrix}
	0.547  \\
	0.619   \\
	0.304 
\end{bmatrix}
+  \begin{bmatrix}
	0.000 & - & - \\
	- & 0.006 & - \\
	- & - & 0.453^\dagger
\end{bmatrix} 
\begin{bmatrix}
	\hat p_{1,t-1}  \\
	\hat p_{2,t-1}   \\
	\hat p_{3,t-1}
\end{bmatrix} + 
\begin{bmatrix}
	0.039^* & 0.012 & 0.071^{**}\\
	0.005 & 0.085^{**} & 0.000 \\
	0.000 & 0.000 & 0.042^*
\end{bmatrix}
\begin{bmatrix}
	Y_{1,t-1}  \\
	Y_{2,t-1}   \\
	Y_{3,t-1}
\end{bmatrix}
\end{equation}


\begin{figure}[h]
\begin{center}
	\includegraphics[width=0.7\linewidth]{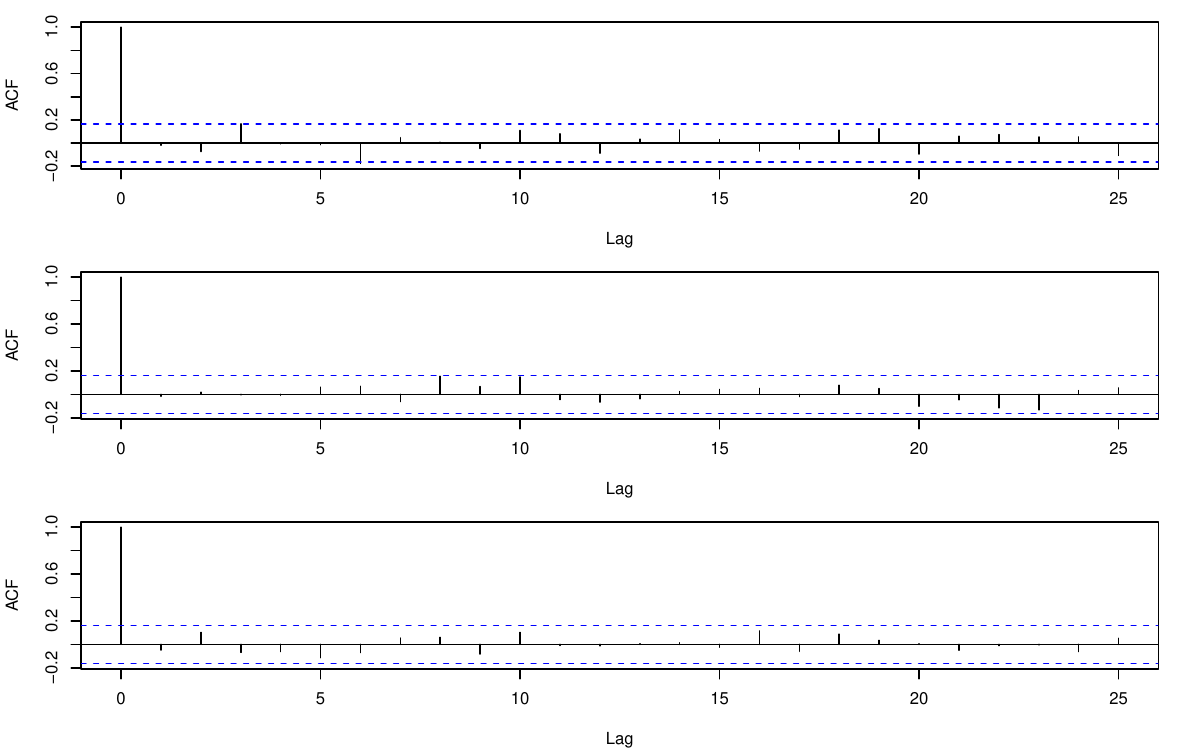}
	\caption{ACF plot of Pearson's residuals for model \eqref{BerWLS estimates}.}%
	\label{acf res}
\end{center}
\end{figure}

\begin{figure}[h]
\begin{center}
	\includegraphics[width=0.7\linewidth]{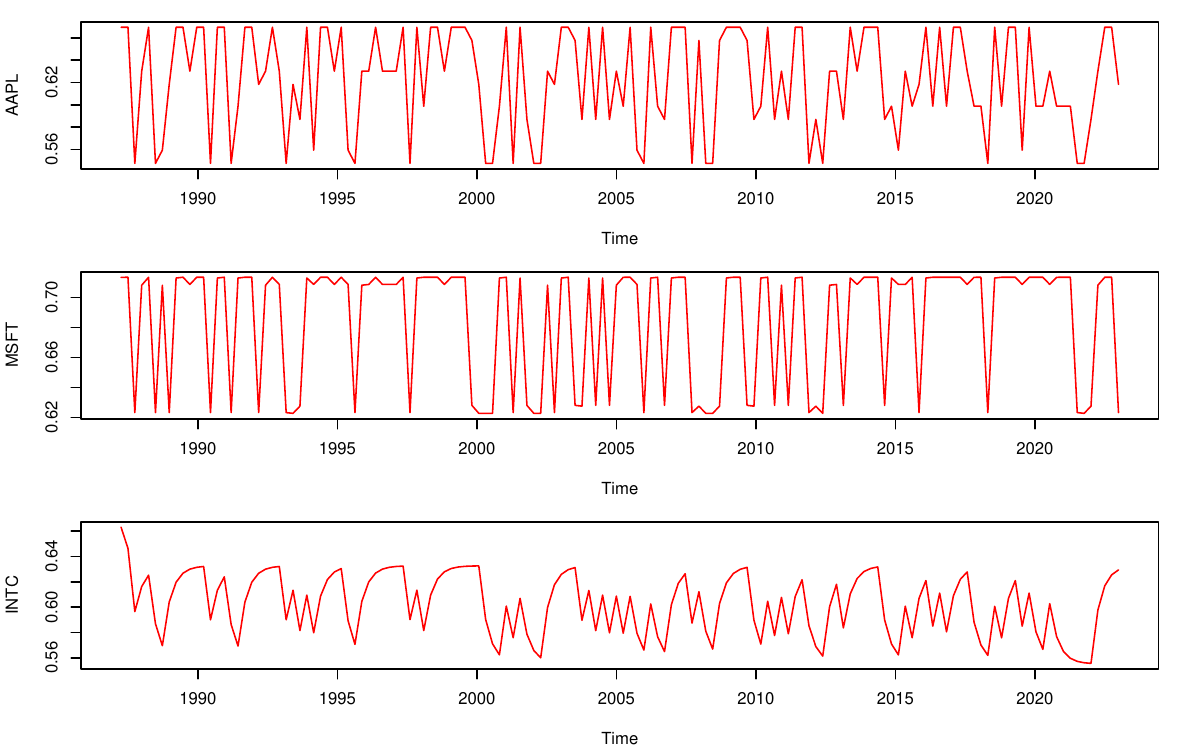}
	\caption{Estimated probability of a positive return using model \eqref{BerWLS estimates} for three main IT companies: Apple (AAPL), Microsoft (MSFT) and Intel (INTC).}%
	\label{est prob}
\end{center}
\end{figure}

Standard errors are estimated using the sample analogues of the asymptotic covariance matrices. The two models deliver similar estimated coefficients but the significance of MWLSE parameters is higher due to the overall reduction of standard error with respect to QMLE providing an indication of improved estimation efficiency. Figure~\ref{acf res} depicts the autocorrelation function (ACF) of the Pearson residuals $( Y_{i,t}-\hat p_{i,t} ) / [\hat p_{i,t}(1 - \hat p_{i,t} )]$. The absence of signal in the residuals indicates a successful model fit. The estimated probabilities $\hat p_{i,t}$ are plotted in Figure~\ref{est prob}. It can be seen that the probability of a positive return is persistently reduced in all the mentioned financial crises or geopolitical instability periods. This pattern is especially pronounced in the MSFT series.

\begin{table}[h!]
\centering
\caption{Mean absolute errors of model \eqref{Bernoulli model} estimated with Bernoulli QMLE (QMLE), first-stage Least Squares Estimator (LSE) and two-stage MWLSE. Lowest values in bold.}
\scalebox{0.95}{
	\begin{tabular}{ccccc}
		\hline
		&	&  Stock & &   \\
		\cline{2-4}
		Model & AAPL & MSFT & INTC & Overall \\ 
		\noalign{\smallskip}\hline \noalign{\smallskip}
		QMLE & 0.4676 & 0.4281 & 0.4774 & 0.4577 \\ 
		LSE & 0.4675 & 0.4280 & 0.4774 & 0.4576 \\ 
		MWLSE & \textbf{0.4669} & \textbf{0.4279} & \textbf{0.4771} & \textbf{0.4573} \\ 
		\hline
	\end{tabular}
}
\label{Tab: mae}
\end{table}
Finally, the prediction with the pool of employed models is evaluated by computing the mean absolute error 
for each company and for the whole multivariate time series. Table 1 presents the results. In all scenarios, the proposed two-stage MWLSE demonstrates superior goodness of fit with respect to the competing quasi-maximum likelihood approaches providing additional evidence of superior performance of two-stage MWLSEs.

\begin{rem} \rm
(Categorical data) The estimation of model \eqref{Bernoulli model} can be extended to more general categorical data with $M$ categories. 
Consider the categorical multivariate random variable $\tilde \Y_t = \left\lbrace 1,2,\dots,M\right\rbrace ^d$ for $t=1,\dots,T$. The $t$-th observation of the categorical
time series can be expressed by the $dq$-dimensional vector $\Y_t = ( \Y^\prime_{(1)t},\dots, \Y^\prime_{(q)t})^\prime$ where $q=M-1$ and, for $j=1,\dots,q$, $\Y_{(j)t}$, are $d$-dimensional binary multivariate time series whose single elements take value $Y_{(j)i,t}=1$ if category $j$ occurs for observation $i$ at time $t$ and $Y_{(j)i,t}=0$ otherwise, for $i=1,\dots,d$. The last category is then defined as $\Y_{(M)t} = 1 - \sum_{j=1}^{q}\Y_{(j)t}$. Analogously, the estimation of \eqref{Bernoulli model} can be carried out on the  aggregated probability vector $\pb_t = ( \pb^\prime_{(1)t},\dots, \pb^\prime_{(q)t})^\prime$ where $\pb_{(j)t}$ is the multivariate $d$-dimensional vector of probabilities associated to category $j$ whose single element is defined by $p_{(j)i,t} = \P(Y_{(j)i,t}=1|\Fb_{t-1})$. To reduce parameters proliferation a parsimonious block diagonal structure specification may be considered for parameter matrices $\A$ and $\B$ with $d \times d$ block diagonal elements $\A_{(1)},\dots, \A_{(q)}$ and $\B_{(1)},\dots, \B_{(q)}$. Finally, the simple constraint $ \vertiii{ \sum_{j=1}^{q} ( \mathbf{C}_{(j)} + \A_{(j)} + \B_{(j)} ) }_\infty <1 $ where $\mathbf{C}_{(j)} = \text{diag}(\mathbf{c}_{(j)})$ allows retaining the summability condition  $\sum_{j=1}^{q} p_{(j)i,t}< 1$ for any $i$ and $t$ and $\pb_{(M)t} = 1 - \sum_{j=1}^{q} \pb_{(j)t}$. Therefore, the two-stage MWLSE introduced in the present contribution also directly applies to categorical data.
\end{rem}

\section{Conclusions}
\label{SEC conclusions}

A two-stage Multivariate Weighted Least Squares Estimator has been proposed for efficient estimation of the conditional mean in multivariate non-negative time series models. The two-stage MWLSE improves on standard QMLE by flexibly accounting for cross-sectional dependence and allowing for general forms of the working conditional covariance matrix, without requiring full specification of the joint distribution of the data. Consistency, asymptotic normality and efficiency of the MWLSE have been established under both general cases and low-level conditions for specific models of interest. 
Simulation results demonstrate substantial efficiency gains of MWLSE over QMLE, particularly in the presence of strong contemporaneous dependence. The empirical application to binary stock return series further illustrates the practical relevance of the approach for disparate non-negative data domains.

Given its computational simplicity and robustness, MWLSE provides a useful alternative to full likelihood methods for a wide range of applied problems, e.g. when maximum likelihood estimation is computationally infeasible or theoretically restrictive.

The proposed weighting sequence structures for the second-stage MWLSE are reasonable and have been shown to be effective in improving the efficiency of estimators in both numerical  and empirical applications. Moreover, a model selection with different competing second-stage MWLSEs based on information criteria can provide further guidance on the selection of the working covariance matrix. Inspired by the quasi-likelihood loss function introduced by \cite{francq_2021two}, future work could focus on the definition of automated data-driven selection processes for working covariance structures.


\section*{Appendix A}
\label{Appendix}
\renewcommand{\theequation}{A.\arabic{equation}}
\renewcommand{\thesection}{A}
\setcounter{equation}{0}
\setcounter{subsection}{0}

\medskip


\subsection{Proofs of the results}
\label{proofs}

\subsubsection*{Proof of Theorem~\ref{Thm. first stage}}
Let $L(\thetab, \W) = \E[l_t(\thetab, \W_t)]$ be the limit log-quasi-likelihood. In what follows the following intermediate results are shown.

\begin{enumerate}[label=(\roman*)] 
\item$\lnorm{L_T(\thetab,\W) - L(\thetab, \W)}_\Theta \to 0$ almost surely, as $T \to \infty$. \label{Cond. uniform convergence}

\item The true parameter value $\thetab_0$ is the unique minimizer of $L(\thetab,\W)$, i.e. \break $\E\left[ l_t(\thetab, \W_t) \right]   > \E\left[  l_t(\thetab_0, \W_t) \right] $ for all $\thetab \in \Theta, \thetab \neq \thetab_0 $. \label{Cond. ineqaulity}

\end{enumerate}

By assumption \ref{Ass. stationarity} the log-quasi-likelihood contribution $l_t(\thetab, \W_t)$ is stationary and ergodic. Moreover, it is uniformly bounded by assumption \ref{Ass. uniform moment}. For the continuity of the quasi-likelihood and the compactness of $\Theta$, \citet[Thm.~2.7]{straumann2006} applies providing the uniform convergence. This concludes the proof of \ref{Cond. uniform convergence}.

Condition~\ref{Cond. ineqaulity} is now proved. First note that by the uniform limit theorem $L(\thetab, \W)= \E[l_t(\thetab, \W_t)]$ is a continuous function and it attains at least a minimum in $\Theta$ since by \ref{Ass. uniform moment} $\Theta$ is compact. It is now proved that such a minimum is unique so that it can be univocally identified. For all $t \in \Z$,  assumption \ref{Ass. pos def.} guarantees that  $Q_t(\lambdab_t)=\E [(\Y_t - \lambdab_t)^\prime \W_t (\Y_t - \lambdab_t) | \Fb_{t-1}]$ is strictly convex for all $\lambdab_t \in \Lambdab$. Therefore, it admits a unique minimum which is $\lambdab_t=\lambdab_t(\thetab_0)$ belonging to the interior of $\Lambdab$, i.e. the solution of the first-order condition
\begin{equation*}
	\frac{\partial Q_t(\lambdab_t)}{\partial \lambdab_t} = 2 \W_t \lambdab_t - 2 \W_t \lambdab_t(\thetab_0) = \zerob\,.
\end{equation*} 
Hence,
\begin{align*}
\E \left[ l_t(\thetab, \W_t) - l_t  (\thetab_0, \W_t) \right] &= \E \left[ \eb_t(\thetab)^\prime \W_t \eb_t(\thetab) \right]  - \E\left[   \eb_t(\thetab_0)^\prime \W_t \eb_t(\thetab_0) \right] \geq 0
\end{align*}
where $\eb_t(\thetab) = \Y_t - \lambdab_t(\thetab) $ and with equality if and only if a.s. $\lambdab_t(\thetab) = \lambdab_t(\thetab_0)$, i.e. if and only if $\thetab = \thetab_0$ by assumption \ref{Ass. identification}.  This concludes the proof of \ref{Cond. ineqaulity}.
The consistency of the estimator $\hat \thetab_1 $ follows from \ref{Cond. uniform convergence}, \ref{Cond. ineqaulity} and the compactness of $\Theta$ by \citet[Lemma~3.1]{PoetscherandPrucha(1997)}. This implies \eqref{consistency}.

Let $\Sb_T(\thetab, \W)$ defined as in \eqref{score mwlse} by replacing $\Vb^*_t(\hat \taub)^{-1}$ with $\W_t$. Then, the single summand of $\Sb_T(\thetab, \W)$ is defined as $\ssb_t(\thetab, \W_t) = \partial\lambdab_{t}(\thetab)'/\partial\thetab \W_t(\Y_t-\lambdab_{t}(\thetab))$. 
Moreover, set $\Hb_T(\thetab, \W) = T^{-1} \sum_{t=1}^{T} -\partial^2 l_t(\thetab, \W_t)/\partial \thetab \partial \thetab^\prime$. To prove the asymptotic normality of the estimator additional intermediate results are established. 

\begin{enumerate}[label=(\alph*)] 
\item $\sqrt{T}  \Sb_T(\thetab_0, \W) \xrightarrow{d} N\left(\zerob, \G(\thetab_0, \W) \right) $  as $T \to \infty$. \label{Cond. clt}
\item  $ \vertiii{ \Hb_T(\thetab, \W) - \Hb(\thetab, \W) }_\Theta \to 0$ almost surely as $T \to \infty$. \label{Cond. uniform convergence hessian}
\end{enumerate}

To prove \ref{Cond. clt} note that $\E\left( \ssb_t(\thetab_0, \W_t) | \Fb_{t-1} \right) = \zerob$. For any $\etab \in \R^m$ with $\etab \neq \zerob$, $\sqrt{T} \etab^\prime \Sb_T(\thetab_0, \W)= T^{-1/2} \sum_{t=1}^{T} U_t$ where $ U_t = \etab^\prime \ssb_t(\thetab_0, \W_t)$. Note that $\left\lbrace U_t, \Fb_t \right\rbrace $ is a stationary martingale difference, and due to \ref{Ass. hessians} it has finite second moments. Then \ref{Cond. clt} follows from the central limit theorem for martingales \cite[]{billingsley1961cltmds} and the Cramér-Wold device.
Note that
\begin{equation} \label{hessian a b}
\frac{\partial^2 l_t(\thetab, \W_t)}{\partial \theta_k \partial \theta_l} = -\frac{\partial \lambdab_t(\thetab)' }{\partial \theta_k} \W_t  \frac{\partial \lambdab_t(\thetab) }{\partial \theta_l} + \frac{\partial^2 \lambdab_t(\thetab)'}{\partial \theta_k \partial \theta_l} \W_t  \left( \Y_t - \lambdab_t(\thetab) \right) =  A_{kl,t}(\thetab, \W_t)+B_{kl,t}(\thetab, \W_t)
\end{equation}
By assumption  \ref{Ass. moment second derivative} the second derivative is a continuous, stationary and ergodic sequence with uniformly bounded expectation. Then, an application of \citet[Thm.~2.7]{straumann2006} provides the condition \ref{Cond. uniform convergence hessian}. 

For $T$ large enough $\hat \thetab_1 \in \dot{\Theta}$ by \ref{Ass. interior}, so the following derivatives exist almost surely
\begin{align*}
\zerob =  \sqrt{T} \Sb_T(\hat\thetab_1, \W) = \sqrt{T} \Sb_T(\thetab_0, \W) - \Hb_T(\bar \thetab, \W)\sqrt{T}( \hat \thetab_1 - \thetab_0) + o_p(1),
\end{align*}
where the first equality comes from the definition \eqref{first-stage estimator} and the second equality is obtained by Taylor expansion at $\thetab_0$ with $\bar \thetab$ lying between $\hat \thetab_1$ and $ \thetab_0$. Therefore, the uniform convergence \ref{Cond. uniform convergence hessian} together with $|| \bar \thetab - \thetab_0|| \leq || \hat \thetab_1 - \thetab_0|| \to 0$ a.s. implies that $\Hb_T(\bar \thetab, \W) \to \Hb(\thetab_0, \W)$ with probability 1. This fact, condition \ref{Cond. clt} and the invertibility of the Hessian matrix in \ref{Ass. hessians} establish the asymptotic normality of the estimator \eqref{asymp. normality}.  \hfill$\square$ 

\subsubsection*{Proof of Theorem~\ref{Thm. second stage}}

Let $\w_t = \text{vec}(\W_t)$ and $\hat \w_t = \text{vec}(\hat \W_t)$. Note that by the multivariate mean value theorem
\begin{equation}
\label{weight consistency}
\lnorm{\hat \w_t - \w_t} = \lnorm{ \kb_t(\hat \taub) - \kb_t(\taub_0)} \leq \vertiii{\frac{\partial \kb_t}{\partial \taub}}_\Tb \lnorm{\hat \taub - \taub_0} 
\end{equation}
which exists by assumption \ref{Ass. weights}. Moreover, the assumption \ref{Ass. moments consistency second stage} and \ref{Ass. stationarity}-\ref{Ass. identification} yield $\lnorm{\hat \taub - \taub_0}  = o(1)$ a.s. as $T \to \infty $. In order to prove the consistency of \eqref{second stage mwlse} the following condition should be verified.

\begin{enumerate}[label={(iii)}]
\item $\lnorm{L_T(\thetab,\hat \W) - L_T(\thetab, \W)}_\Theta \to 0$ almost surely, as $T \to \infty$.
\label{Cond. uniform convergence second stage}
\end{enumerate}
Consider the norm related to the single summand of the weighted least squares criterion
\begin{align*}
\lnorm{l_t(\thetab,\hat \W_t) - l_t(\thetab, \W_t)}_\Theta &
  \leq \vertiii{\hat \W_t - \W_t} \lnorm{\eb_t}_\Theta^2 
\leq  \lnorm{\hat \w_t - \w_t}  \lnorm{\eb_t}_\Theta^2  \\
& \leq \lnorm{\hat \taub - \taub_0} \vertiii{\frac{\partial \kb_t}{\partial \taub}}_\Tb \lnorm{\eb_t}_\Theta^2  \\
& = o(1) \quad a.s.
\end{align*}
where 
 the second inequality comes from $\vertiii{\X} \leq \vertiii{\X}_F = \lnorm{\text{vec}(\X)}$ for any square matrix $\X$, and  the third inequality comes from \eqref{weight consistency}. Finally, the last equality is satisfied  by \ref{Ass. weights} and the ergodic theorem. 
An application of Cesaro's lemma proves \ref{Cond. uniform convergence second stage} as $T \to \infty$,
\begin{equation*}
\lnorm{L_T(\thetab,\hat \W) - L_T(\thetab, \W)}_\Theta \leq	T^{-1}\sum_{t=1}^{T} \lnorm{l_t(\thetab,\hat \W_t) - l_t(\thetab, \W_t)}_\Theta \xrightarrow{a.s.} 0\,.
\end{equation*}
Therefore, the consistency \eqref{consistency second stage} follows as in the proof of Theorem~\ref{Thm. first stage}. 
Define $\Sb_T(\thetab_0, \W(\taub))=T^{-1} \sum_{t=1}^{T} \ssb_T(\thetab_0, \W_t(\taub)) $ where $\W_t(\taub)=\Vb_t^*(\taub)^{-1}$  and $\Jb_T(\taub) = \partial \Sb_T(\thetab_0, \W(\taub))/ \partial \taub$. The asymptotic normality of the two-stage estimator follows from the condition stated below.
\begin{enumerate}[label={(c)}]
\item \label{Cond. uniform convergence hessian second stage}  $ \vertiii{\Hb_T(\thetab, \hat \W) - \Hb_T(\thetab, \W)}_\Theta \to 0$ almost surely as $T \to \infty$.
\end{enumerate}
\begin{enumerate}[label={(d)}]
\item \label{Cond. uniform convergence hessian second stage 2} $ \vertiii{\Jb_T }_\Tb \to 0$ almost surely as $T \to \infty$.
\end{enumerate}
Define $\Hb_T(\thetab, \W(\taub)) = T^{-1} \sum_{t=1}^{T} -\partial^2 l_t(\thetab, \W_t(\taub))/\partial \thetab \partial \thetab^\prime$ and that from \eqref{hessian a b} its $(k,l)$-element  is
\begin{equation*}
	H_{kl,T}(\thetab, \W(\taub)) = - A_{kl,T}(\thetab, \W(\taub)) - B_{kl,T}(\thetab, \W(\taub))
\end{equation*}
where $A_{kl,T}(\thetab, \W(\taub)) = T^{-1} \sum_{t=1}^{T} A_{kl,t}(\thetab, \W_t(\taub))$ and $B_{kl,T}(\thetab, \W(\taub)) = T^{-1} \sum_{t=1}^{T} B_{kl,t}(\thetab, \W_t(\taub))$.
Standard algebra for $\etab,\nub \in \R^d$ and $\X \in \R^{d \times d}$ provides $\etab'\X \nub = \text{tr}(\X \nub \etab') = \vecc(\X)'\vecc(\nub \etab')$. Hence, for some $\bar \taub$ between $\hat \taub$ and $\taub_0$ 
\begin{align*}
\frac{\partial A_{kl,T}( \W(\bar \taub))}{\partial \taub} = \lnorm{\frac{\partial A_{kl,T}(\thetab, \W(\bar \taub))}{\partial \taub}}_\Theta &\leq \frac{1}{T} \sum_{t=1}^{T} \vertiii{\frac{\partial \vecc(\W_t(\bar \taub))}{\partial \taub}}\lnorm{\vecc\left(  \frac{\partial \lambdab_t }{\partial \theta_l}   \frac{\partial \lambdab_t' }{\partial \theta_k} \right) }_\Theta \\
&= \frac{1}{T} \sum_{t=1}^{T} \vertiii{\frac{\partial \kb_t(\bar \taub)}{\partial \taub}} \, \vertiii{\frac{\partial \lambdab_t }{\partial \theta_l}   \frac{\partial \lambdab_t' }{\partial \theta_k}}_{F,\Theta}\,.
\end{align*}
  A Taylor expansion in $\taub$ provides
\begin{align*}
\lnorm{A_{kl,T}(\thetab, \hat \W) - A_{kl,T}(\thetab, \W)}_\Theta \leq \lnorm{\frac{\partial A_{kl,T}( \W)}{\partial \taub}}_\Tb \lnorm{\hat \taub - \taub_0} \to 0 ~ a.s.
\end{align*}
by \ref{Ass. hessian second stage} and  the ergodic theorem. Similar arguments entail $\lnorm{B_{kl,T}(\thetab, \hat \W) - B_{kl,T}(\thetab, \W)}_\Theta  \to 0$ a.s. from the second moment  in \ref{Ass. hessian second stage}. This proves condition \ref{Cond. uniform convergence hessian second stage}.

The $k$-th row of the single element of $\Jb_T(\taub)$ is   $\Jb_{k,t}(\taub) = \partial/\partial \taub \left[  \partial \lambdab_t(\thetab_0)'/\partial \theta_k  \W_t(\taub)  \left( \Y_t - \lambdab_t(\thetab_0) \right) \right] $ where $\E(\Jb_{k,t}(\taub))=0$ for every $k=1,\dots,m$ and by analogous arguments above
$$
\lnorm{\Jb_{k,t}}_\Tb \leq \vertiii{\frac{\partial \kb_t}{\partial \taub}}_\Tb  \vertiii{\left( \Y_t - \lambdab_t(\thetab_0) \right) \frac{\partial \lambdab_t(\thetab_0)'}{\partial \theta_k}  }_F
$$
which has a finite expectation by \ref{Ass. hessian second stage}. Moreover, since $\Jb_t(\taub)$ is stationary ergodic and is continuous on the compact set $\Tb$,  \citet[Thm.~2.7]{straumann2006} applies providing \ref{Cond. uniform convergence hessian second stage 2}.

Analogously to the proof of Theorem~\ref{Thm. first stage}, a Taylor expansion around $\thetab_0$ provides almost surely
\begin{align*}
	\zerob =  \sqrt{T} \Sb_T(\hat\thetab, \hat \W) = \sqrt{T} \Sb_T(\thetab_0, \hat \W) - \Hb_T(\tilde \thetab, \hat \W)\sqrt{T}( \hat \thetab - \thetab_0) + o_p(1),
\end{align*} 
where $||\tilde \thetab - \thetab_0 || \leq  ||\hat \thetab - \thetab_0 ||$ and $\Hb_T(\tilde \thetab, \hat \W) \to \Hb(\thetab_0, \W)$ a.s. by an application of \ref{Cond. uniform convergence hessian} plus \ref{Cond. uniform convergence hessian second stage}.
It is left to show that $\sqrt{T} \Sb_T( \thetab_0, \hat \W) = \sqrt{T} \Sb_T(\thetab_0, \W) + o_p(1)$ which is proved by further applying Taylor's expansion theorem around $\taub_0$ 
\begin{align*}
	\sqrt{T} \Sb_T(\thetab_0, \hat \W) = \sqrt{T} \Sb_T(\thetab_0,  \W) - \Jb_T(\bar \taub)\sqrt{T}( \hat \taub - \taub_0) + o_p(1),
\end{align*} 
and \ref{Cond. uniform convergence hessian second stage 2} plus \ref{Ass. bounded in prob}.
The asymptotic normality \eqref{asymp. normality second stage} follows from analogous arguments as in the proof of Theorem~\ref{Thm. first stage}.

Finally, in the case of correctly specified conditional variance, the weighting sequence is $\W_t=\Vb_t^{-1}$. Therefore, $\Hb(\thetab_0,\W)=\G(\thetab_0,\W)$ and $\Sigmab=\G^{-1}$. Moreover,
\begin{equation*}
	\Sigmab - \G^{-1} = \text{Var} \Big[ \Hb^{-1}(\thetab_0, \W) \ssb_t(\thetab_0, \W_t) - \G^{-1} \ssb_t(\thetab_0, \Vb^{-1}_t)  \Big] 
\end{equation*}
being necessarily positive semi-definite. \hfill$\square$

\subsubsection*{Proof of Theorem~\ref{Thm. efficiency ef}}
Define the notation $\nabla \etab(\lambdab) = \partial \etab(\lambdab)/ \partial \lambdab$, $\nabla_i \etab(\lambdab) = \partial \etab(\lambdab)/ \partial \lambda_i$, and $\nabla_{ij} \etab(\lambdab) = \partial \etab(\lambdab)/ \partial \lambda_i \partial \lambda_j$. An analogous notation applies to $a(\lambdab)$.
It is now proved that a random vector $\Y$ with mean $\lambdab$, covariance matrix $\Vb$ and density $f_{\lambdab}$ belonging to a regular multivariate exponential family satisfies
	\[
\nabla a(\lambdab) = \nabla \etab(\lambdab)'\lambdab\,, \quad \quad \nabla \etab(\lambdab)' = \nabla \etab(\lambdab)' \Vb \nabla \etab(\lambdab) \,.
	\]
The first equality is satisfied since
	\[
	0 = \frac{\partial}{\partial \lambda_i} \int_\Yc f_{\lambdab}(\yb) \, d\mu(\yb) = \int_\Yc h(\yb) e^{\etab(\lambdab)'\yb - a(\lambdab)} \left[ \nabla_i \etab(\lambdab)'\yb - \nabla_i a(\lambdab) \right] \, d\mu(\yb) =  \nabla_i \etab(\lambdab)'\lambdab - \nabla_i a(\lambdab)\,,
	\]
whereas the second equality comes from
	\begin{align*}
	0 = \frac{\partial^2}{\partial \lambda_i \partial \lambda_j } \int_\Yc f_{\lambdab}(\yb) \, d\mu(\yb) = &\int_\Yc h(\yb) e^{\etab(\lambdab)'\yb - a(\lambdab)} \left[ \nabla_i \etab(\lambdab)'\yb - \nabla_i a(\lambdab) \right] \left[ \nabla_j \etab(\lambdab)'\yb - \nabla_j a(\lambdab) \right] \, d\mu(\yb) \\  & + \int_\Yc h(\yb) e^{\etab(\lambdab)'\yb - a(\lambdab)} \left[  \nabla_{ij} \etab(\lambdab)'\yb - \nabla_{ij} a(\lambdab) \right] \, d\mu(\yb)\,,
	\end{align*}
	and therefore
\begin{align*}
	0 &= \int_\Yc f_{\lambdab}(\yb) \nabla_i \etab(\lambdab)' \left( \yb -\lambdab \right) \left( \yb - \lambdab \right)' \nabla_j \etab(\lambdab) \, d\mu(\yb) +   \nabla_{ij} \etab(\lambdab)'\lambdab - \nabla_{ij} a(\lambdab)\\
	&= \nabla_i \etab(\lambdab)' \Vb \nabla_j \etab(\lambdab) + \nabla_{ij} \etab(\lambdab)'\lambdab - \nabla_{ij} a(\lambdab) \,.
\end{align*}
By the product rule it also holds that
$
\nabla_{ij} a(\lambdab) =  \nabla_{ij} \etab(\lambdab)'\lambdab +  \partial \etab_j(\lambdab)/ \partial \lambda_i
$.
	It follows that
	\[
\frac{\partial \etab_j(\lambdab)}{ \partial \lambda_i} = \nabla_i \etab(\lambdab)' \Vb \nabla_j \etab(\lambdab)\,
	\]
	being the $(i,j)$-element of the matrix $\nabla \etab(\lambdab)'$. Hence $\nabla \etab(\lambdab)$ is symmetric and by assumption invertible, yielding $\nabla \etab(\lambdab) =\nabla \etab(\lambdab)\Vb \nabla \etab(\lambdab)$ or equivalently, 
	$$
	\nabla \etab(\lambdab) = \Vb ^{-1}\,.
	$$
	Note that for $t\in\Z$ the function $ M_t(\lambdab_t) = \E \left[ \etab(\lambdab_t)'\Y_t - a(\lambdab_t) | \Fb_{t-1}\right]$ is strictly concave in $\lambdab_t \in \Lambdab$ since 
	\begin{equation*}
		\frac{\partial M_t(\lambdab_t)}{\partial \lambdab_t } = \nabla \etab(\lambdab_t)'\lambdab_t(\thetab_0) - \nabla a(\lambdab_t) = \nabla \etab(\lambdab_t)'\left( \lambdab_t(\thetab_0) - \lambdab_t \right) = \Vb_t^{-1} \left( \lambdab_t(\thetab_0) - \lambdab_t \right)\,,
	\end{equation*}
	\begin{equation*}
		 \quad \quad \frac{\partial^2 M_t(\lambdab_t)}{\partial \lambdab_t \partial \lambdab_t'} = - \Vb_t^{-1} < 0\,.
	\end{equation*}
	Moreover, the unique maximizer which satisfies the first-order condition is $\lambdab_t =\lambdab_t(\thetab_0)$. Therefore, the conditional log-likelihood of $\Y_t$ given $\Fb_{t-1}$
	\[
	\ell_t(\thetab) = \log f_{\lambdab_t(\thetab)}(\Y_t) = \etab \left( \lambdab_t(\thetab) \right)' \Y_t - a \left( \lambdab_t(\thetab) \right)
	\]
	is such that $\E[\ell_t(\thetab)] \leq \E[\ell_t(\thetab_0)]$ with equality if and only if a.s. $\lambdab_t(\thetab) = \lambdab_t(\thetab_0)$. This fact and the same arguments of the proof of Theorem~\ref{Thm. first stage} entail that the MLE of $\thetab_0$, say $\hat \thetab_{ML}$,
satisfies
	\[
	\sqrt{T} \left( \hat{\thetab}_{ML} - \thetab_0 \right) \xrightarrow{d} N \left( \zerob, \G^{-1} \right)\,, \quad \quad \G = \E \left[ \frac{\partial\lambdab_{t}(\thetab_0)'}{\partial\thetab} \Vb_t^{-1} \frac{\partial\lambdab_{t}(\thetab_0)}{\partial\thetab} \right]
	\]
	as $T \to \infty$, where $\G$ is positive definite since $\forall \nub \in \R^m$, $\nub \neq \zerob$ 
	$$
 \nub'\G\nub= \E \left[ \nub' \frac{\partial\lambdab_{t}(\thetab_0)'}{\partial\thetab} \Vb_t^{-1} \frac{\partial\lambdab_{t}(\thetab_0)}{\partial\thetab} \nub \right] = \E \left[ \pib'_t  \Vb_t^{-1} \pib_t  \right] > 0\,
	$$
	and $ \pib_t = \partial\lambdab_{t}(\thetab_0)/\partial\thetab \nub \neq \zerob $.
	The proof concludes by noting that $\G$ is the inverse of the asymptotic variance of the two-stage MWLSE, as defined in equation~\eqref{asymp. normality second stage eff}. \hfill$\square$\\
	
	\subsubsection*{Proof of Proposition~\ref{Prop. mult ingarch}}
	
The result of the proposition follows by proving the assumptions of Theorem~\ref{Thm. first stage}. \citet[Thm.~2]{tru2021} guarantees that the count process in \eqref{ingarch} is stationary and ergodic. Moreover, all the moments of $\Y_t$ exist. Following the arguments of \citet[Supp. mat.~S2.11]{armillotta2025copula}, this implies that  $\E\lnorm{\lambdab_t}^p_\Theta < \infty$, $ \E \lnorm{ \partial \lambdab_t/\partial \theta_k }^p_\Theta < \infty$, and $ \E \lnorm{ \partial^2 \lambdab_t/\partial \theta_k \partial \theta_l }^p_\Theta < \infty$ for every $p \geq 1$ and every $k,l=1,\dots,m$.
Therefore, $\E \lnorm{l_t(\thetab, \W_t)}_\Theta  \leq \underline{\lambda}^{-1} \E\lnorm{\Y_t-\lambdab_t}^2_\Theta   < \infty$ and by Cauchy-Schwarz inequality the moments in \ref{Ass. moment second derivative} are finite. For the same reasons, $\E \, \vertiii{\Vb_t}^2 < \infty$ entailing that the matrices in \ref{Ass. hessians} exist and are finite. 
 Condition~\ref{Ass. identification} follows from $|\B_0|_e 1_d \nsim 0$ and arguments analogous to \citet[Supp. mat.~S2.3]{armillotta2025copula} where $\lambdab_t \succeq \textbf{c}$ lies in the interior of $\Lambdab=(0,+\infty)^d$. Finally, following  \citet[Supp. mat.~S2.6]{armillotta2025copula} the columns of  $\partial\lambdab_{t}(\thetab_0)/\partial\thetab$  are linearly independent. According to the arguments at the end of the proof of Theorem~\ref{Thm. efficiency ef}, this fact guarantees that the matrices in \ref{Ass. hessians} are positive definite. \qed
 
 	\subsubsection*{Proof of Proposition~\ref{Prop. mult ingarch second stage}}
 	
 	The result of the proposition follows by proving the assumptions of Theorem~\ref{Thm. second stage}.
 	In this example there are no nuisance parameters therefore $\hat \taub = \hat \thetab_1$ so \ref{Ass. moments consistency second stage} and \ref{Ass. bounded in prob} apply.
 Let $\underline c =\min_{i=1,\dots,d} c_i$ where $c_i$ is the single element of the intercept vector $\textbf{c}$. 
 	Note that the eigenvalues of $\Pb^*(\thetab)$ are $ 1 + (d-1) r^*(\thetab)$ with multiplicity 1 and $\ 1- r^*(\thetab)$ with multiplicity $d-1$. Therefore, by condition \ref{pos def bound equi} it holds that $\lambda_{\min}(\Pb^*(\thetab)) \geq \underline{\lambda}_P > 0$ which satisfies \ref{pos def bound} by recalling that $\lambda_{\min}(\Vb^*_t(\thetab))^{-1}=\vertiii{\Vb^*_t(\thetab)^{-1}} \leq \vertiii{\Db^*_t(\thetab)^{-1/2}} \vertiii{\Pb^*(\thetab)^{-1}} \vertiii{\Db^*_t(\thetab)^{-1/2}} \leq \underline{c}^{-1}\lambda_P^{-1}=\underline{\lambda}^{-1}$.
 
 	The single element of $\partial \kb_t(\thetab)/\partial \theta_k$ can be written has $\partial V^*_{ij,t}(\thetab)^-/\partial \theta_k$ where $V^*_{ij,t}(\thetab)^-$ is the $(i,j)$-element of the inverse pseudo-covariance matrix $\Vb_t^*(\thetab)^{-1}$. 
 	By the structure \eqref{optimal weights} and the chain rule
 	\begin{equation*}
 		\left| \frac{\partial V^*_{ij,t}(\thetab)^-}{\partial \theta_k}\right| \leq \frac{1}{2 \underline{c}^2} \left( \left| \frac{\partial \lambda_{i,t}(\thetab)}{\partial \theta_k}\right| +\left| \frac{\partial \lambda_{j,t}(\thetab)}{\partial \theta_k}\right| \right) \left| r^-_{ij}(\thetab)\right| + \frac{1}{c} 	\left| \frac{\partial  r^-_{ij}(\thetab)}{\partial \theta_k}\right|\,,
 	\end{equation*} 
where $ r^-_{ij}(\thetab)$ is the single element of the matrix $\Pb^*(\thetab)^{-1}$. By the equicorrelation structure it follows that $\Pb^*(\thetab)^{-1}=b(\thetab) \Jb_d + [a(\thetab)-b(\thetab)]\I_d$, with $a(\thetab)=[1+(d-2)r^*(\thetab)]/D(\thetab)$ and $b(\thetab)=-r^*(\thetab)/D(\thetab)$ where $D(\thetab)=\left\lbrace [1-r^*(\thetab)][1+(d-1)r^*(\thetab)]\right\rbrace$. For a proof see \citet[p.~67]{rao_2002}. Hence, $r^-_{ii}(\thetab)=a(\thetab)$ and $r^-_{ij}(\thetab)=b(\thetab)$ for $i \neq j$. In absolute values, the numerators of these quantities are upper-bounded by positive constants whereas, by \ref{pos def bound equi}, the denominator is lower-bounded by positive constants therefore $| r^-_{ij}(\thetab)|$ is bounded. Recall that $|\partial  r^-_{ij}(\thetab)/\partial \theta_k|=|\partial  r^-_{ij}(\thetab)/\partial r^*(\thetab) | | \partial  r^*(\thetab)/\partial \theta_k |$ where the first (absolute) derivative is bounded by analogous arguments above. The second derivative has a denominator lower-bounded by $\underline{c}^2$ while the numerator is composed of products and sums of $\lambda_{i,t}(\thetab)^p$ and its first derivatives up to $p=2$. Therefore, it has finite moments of any order uniformly over $\Theta$. Finally, iterated applications of Cauchy-Schwarz inequality provide \ref{Ass. weights} and \ref{Ass. hessian second stage}. \qed

\subsubsection*{Proof of Proposition~\ref{Prop. mult acd}}

By \citet[Cor.~1]{debaly_truquet_2023} and \citet[Prop.~A.2]{davis2016}, the positive-valued process defined in \eqref{acd} is stationary and ergodic. Following the arguments of \citet[Supp. mat.~S2.11]{armillotta2025copula},   $\E\lnorm{\lambdab_t}^p_\Theta < \infty$, $ \E \lnorm{ \partial \lambdab_t/\partial \theta_k }^p_\Theta < \infty$, and $ \E \lnorm{ \partial^2 \lambdab_t/\partial \theta_k \partial \theta_l }^p_\Theta < \infty$ for $1 \leq p \leq 4$ and every $k,l=1,\dots,m$.
Hence, the result follows from the same
arguments used in the proofs of Propositions~\ref{Prop. mult ingarch}-\ref{Prop. mult ingarch second stage}. \qed

\section*{Acknowledgements}
The author acknowledges financial support from the EU Horizon Europe programme under the Marie Skłodowska-Curie grant agreement No.101108797.

\section*{Declarations}

\textbf{Conflict of interest} The author states that there is no conflict of interest.

\bibliographystyle{apalike}
\bibliography{poisson_nar,references}

\end{document}